\documentclass[11pt,a4paper]{article}
\pdfoutput=1
\usepackage{jheppub}

\usepackage[usenames,dvipsnames,table]{xcolor}
\usepackage{graphicx,amsmath,amssymb,amsthm,multirow,array,bm,bbm,esint,slashed,hyperref
}
\usepackage{xfrac,faktor}
\usepackage[mathscr]{eucal}
\usepackage{subfig}
\usepackage{comment}
\usepackage{braket}
\usepackage{longtable}

\newcommand\beq{\begin{equation}}
\newcommand\eeq{\end{equation}}
\newcommand{\nn}{\nonumber}

\newcommand{\M}{{\cal M}}

\newcommand{\R}{\mathbb R}

\newcommand{\C}{\mathbb C}
\newcommand{\Z}{\mathbb Z}

\def\be{\begin{equation}}
\def\ee{\end{equation}}
\def\bea{\begin{eqnarray}}
\def\eea{\end{eqnarray}}

\newcommand{\Tr}{{\rm {Tr}}}

\newcommand{\ba}{\begin{eqnarray}}
\newcommand{\ea}{\end{eqnarray}}

\newcommand{\f}{\frac}
\newcommand{\s}{\sqrt}

 \def\vep{\varepsilon}

 \def\f {\frac}

\title{Liouville Theory and the Weil--Petersson Geometry of Moduli Space}
\author[a,b]{Sarah M. Harrison,}
\author[a]{Alexander Maloney,}
\author[c]{and Tokiro Numasawa}

\affiliation[a]{Department of Physics, McGill University, Montreal, QC, Canada}
\affiliation[b]{Department of Mathematics and Statistics, McGill University, Montreal, QC, Canada}
\affiliation[c]{Institute for Solid State Physics, University of Tokyo, Kashiwa 277-8581, Japan}

\emailAdd{sarah.harrison@mcgill.ca}
\emailAdd{alex.maloney@mcgill.ca}
\emailAdd{numasawa@issp.u-tokyo.ac.jp}

\abstract{Liouville theory describes the dynamics of surfaces with constant negative curvature and can be used to study the Weil-Petersson geometry of the moduli space of Riemann surfaces.  
This leads to an efficient algorithm to compute the Weil--Petersson metric to arbitrary accuracy using Zamolodchikov's recursion relation for conformal blocks. 
For example, we compute the metric on $\mathcal M_{0,4}$ numerically to high accuracy by considering Liouville theory on a sphere with four punctures.  We numerically compute the eigenvalues of the Weil-Petersson Laplacian, and find evidence that the obey the statistics of a random matrix in the Gaussian Orthogonal Ensemble.}

\arxivnumber{2210.08098}

\begin{document}
\maketitle

\section{Introduction and Discussion}

Two dimensional Conformal Field Theory (CFT) has proven to be a useful tool in both physics and mathematics, with applications ranging from the study of critical phenomena to string theory and topological quantum field theory.  One of its most intriguing applications is to the geometry of surfaces.  This is most clear in the Liouville conformal field theory introduced by Polyakov \cite{Polyakov:1981rd}, which can be regarded as a quantum theory of geometry in two dimensions.  The classical solutions of this theory are two dimensional surfaces of constant negative curvature, possibly with sources.  
This allows us to use Liouville theory to study the moduli space of Riemann surfaces, which can be regarded as the space of hyperbolic metrics (or alternately the space of conformal structures) on a surface of given topology.  Our goal will be to use conformal field theory methods to study the Weil-Petersson (WP) geometry of moduli space.  Our main result is an efficient algorithm to approximate the Weil-Petersson metric to arbitrary accuracy.  

Although we will restrict our attention to a relatively simple case -- the moduli space ${\cal M}_{0,4}$ of the four-punctured sphere -- this strategy works in general and can be used to study moduli spaces of surfaces both with and without boundary.  
A key role will be played by Zamolodchikov's recursion relations for conformal blocks \cite{Zamolodchikov:1984eqp, zamolodchikov1987conformal}, which leads to an approximation that converges extremely rapidly.  For example, the first non-trivial order in this expansion gives a result for the volume of moduli space that is correct to one part in $10^{-7}$.  
As an application of these results we will compute numerically the first 200 eigenvalues of the Weil-Petersson Laplacian on moduli space.  We will verify that these obey the expected Weyl law governing the asymptotic behaviour of the spectrum of the Laplacian.  We will also study the spacings of the eigenvalues, and show that these exhibit the expected behaviour for a chaotic system in the Gaussian Orthogonal Ensemble (GOE) universality class.

The relationship between Liouville theory and the classical and quantum geometry of surfaces has been discussed previously; early discussions include \cite{Takhtajan:1993vt, Matone:1993tj, Takhtajan:1994vt}.  One major achievement in the study of Liouville theory was the remarkable solution -- described in \cite{Dorn:1994xn, Zamolodchikov:1995aa, Ponsot:1999uf, Ponsot:2000mt, Teschner:2001rv, Teschner:2003at} -- of the constraints of crossing symmetry to determine the three-point coefficients of the theory exactly.  This opened the door to quantitative connections between Liouville theory and the geometry of surfaces.  We particularly highlight the work of \cite{Hadasz:2005gk}, which used the same technique that we will describe below, although to study a different quantity.

One of our central results is that these CFT techniques allow one to approximate the Weil-Petersson metric with sufficient accuracy that one can study the chaotic dynamics on moduli space explicitly. Although our focus is on the spectrum of the WP Laplacian, one could also study the classical dynamics of motion on moduli space.  This subject has been of considerable interest in the mathematics literature, where various chaotic properties have been discussed previously (see e.g.  \cite{2008arXiv0811.2362E, rudnick2022goe}).  The study of quantum chaos also has a long history in the physics literature, where a generic Hamiltonian is expected to exhibit statistical properties which are governed by Random Matrix Theory (RMT) \cite{mehta1991random, PhysRevLett.52.1, refId0}.   The most general ensemble of random matrices is the Gaussian Unitary Ensemble, which describes Hamiltonians without time-reversal symmetry.  The GOE, which is the ensemble relevant for the WP Laplacian, describes Hamiltonians with time-reversal symmetry.  RMT has also made an appearance in the study of gravitational systems, since black holes are expected to exhibit quantum chaos and therefore be described by RMT as well \cite{Cotler:2016fpe}.  There is a substantial literature which studies the RMT behaviour of a variety of quantum mechanical systems, but examples where this can be studied in genuine field theories or gravitational systems are rare.\footnote{See e.g. \cite{Afkhami-Jeddi:2021qkf} for a recent example.}

Our results have potential applications in quantum gravity.
In particular, the Weil-Petersson Laplacian is a natural Hamiltonian acting on a phase space of metrics, so our computation can be regarded as a study of a simple model of quantum gravity based on the quantization of moduli space.  This appears naturally in three dimensional theories of AdS gravity, where in a canonical formulation the phase space of a pure theory of gravity can be constructed from the moduli space of Riemann surfaces \cite{Verlinde:1989ua,Scarinci:2011np, Kim:2015qoa, Maloney:2015ina, Eberhardt:2022wlc}.  The computations in this paper are most simply related to so-called ``chiral gravity" in AdS$_3$ \cite{Li:2008dq}, where the phase space is precisely the moduli space of Riemann surfaces endowed with the Weil-Petersson symplectic structure, as discussed recently in \cite{Maloney:2015ina, Eberhardt:2022wlc}.\footnote{This is to be contrasted with general relativity in AdS$_3$, where the phase space is the cotangent bundle of moduli space.}  It would be interesting to investigate whether, following \cite{Maloney:2015ina}, wave functions of the type we have constructed can be viewed as a class of black hole microstates.  Our computations may also be applicable to two dimensional theories of gravity such as Jackiw-Teitelboim gravity in AdS$_2$, where the gravitational path integral can be written as an integral over the moduli space of Riemann surfaces with the Weil-Petersson volume form \cite{Saad:2019lba}.

In \S \ref{s:review} we  briefly review a few important aspects of Liouville field theory and its relationship with the geometry of surfaces, before moving in \S \ref{s:Modgeom} to a review of necessary features of the moduli space of curves.  \S \ref{sec:WPmetric} contains the main computation, which uses the Zamolodchikov recursion relations to develop an accurate approximation to the Weil-Petersson metric, which is checked in a variety of ways.  \S \ref{s:spectrum} applies this to a numerical computation of the eigenvalues of the Weil-Petersson Laplacian, along with a discussion of their properties.  Appendix \ref{app:eigenvalues} lists the first 100 eigenvalues and includes a brief discussion of the numerical accuracy of these results.

\section{Classical and quantum Liouville theory}\label{s:review}

In this section we review the basics of Liouville quantum field theory and its connection to the geometry of Riemann surfaces. We are brief, and only highlight a few aspects salient for our analysis in the coming sections. For more complete reviews of Liouville QFT, see e.g.\cite{Seiberg:1990eb,Nakayama:2004vk,Teschner:2001rv}. 

\subsection{The Liouville CFT}
The action of Liouville theory on a Riemann surface $\Sigma$ with complex coordinates $(z,\bar z)$ (we use the conventions of \cite{Zamolodchikov:1995aa}) is
\be\label{eq:SL}
S_L[\phi]: = \int_\Sigma d^2z \sqrt g \left ({1\over 4\pi} g^{\alpha\beta} \partial_\alpha \phi\partial_\beta \phi + \mu e^{2b \phi}\right ),
\ee
where $\mu$ is known as the Liouville cosmological constant. If the surface $\Sigma$ has a boundary, this must be augmented by a certain choice of boundary conditions for $\phi$ at $\partial \Sigma$, but this will not be important for our discussion.  
The quantum theory described by this action is a conformal field theory with central charge $c= 1+ 6Q^2$, where $Q= b + 1/b$. The primary operators of this theory are the Liouville vertex operators $V_{\alpha}(z,\bar z)$ with conformal dimension $\Delta_\alpha= \alpha(Q-\alpha)$ where
$$ \alpha = {Q\over 2} + i P, ~~ P \in \R^+.$$
We will refer to the parameter $P$ as the Liouville momentum.

The quantum theory is characterized by the $n$-point correlation functions of primary operators:
\be\label{corr}
\langle V_{\alpha_1}(z_1,\bar z_1) \ldots V_{\alpha_n}(z_n,\bar z_n) \rangle.
\ee
Using operator product expansions these correlation functions can be reduced to expressions involving the three point function coefficients $C(\alpha_i,\alpha_j,\alpha_k)$ 
of primary operators and the Virasoro conformal blocks $\mathcal{F}_P$.
For example, the four point function---which is the crucial ingredient we will use to study the geometry of $\mathcal M_{0,4}$---can be represented as an integral over the Liouville momentum $P$ of an internal operator exchanged in the $s$--channel as,
\be
\langle V_{\alpha_4}(\infty,\infty) V_{\alpha_3}(1,1)V_{\alpha_2}(x,\bar{x}) V_{\alpha_1}(0,0)   \rangle = \int dP C(\alpha_1,\alpha_2, \alpha) C(\alpha_3,\alpha_4,Q- \alpha ) \left|\mathcal{F}_P\begin{bmatrix}\Delta_3 & \Delta_2 \\ \Delta_4  & \Delta_1 \end{bmatrix}(x)\right|^2, \label{eq:4ptLFTq}
\ee
where we have introduced the cross ratio $$x = {(z_1-z_2)(z_3-z_4)\over (z_1-z_4)(z_3-z_2)}.$$
The structure constants $C(\alpha_1,\alpha_2,\alpha_3)$ of Liouville have been determined explicitly by solving the constraints of conformal invariance and crossing symmetry, and are given by the so-called DOZZ formula \cite{Dorn:1994xn,Zamolodchikov:1995aa}.

The 4-point Virasoro conformal block $\mathcal{F}_P\begin{bmatrix}\Delta_3 & \Delta_2 \\ \Delta_4  & \Delta_1 \end{bmatrix}$ captures the contributions to the four point function of an internal primary operator with Liouville momentum $P$ along with all its Virasoro descendants.
Closed form expressions for $\mathcal{F}_P$ exist only in certain special cases, but we can efficiently compute its expansion around $x = 0$ to arbitrary order using the recursion relations of Zamolodchikov \cite{zamolodchikov1987conformal}.  For example, the first two terms in the expansion are
\be\label{eq:FP}
\mathcal{F}_P\begin{bmatrix}\Delta_3 & \Delta_2 \\ \Delta_4  & \Delta_1 \end{bmatrix} = x^{\Delta-\Delta_2- \Delta_1}
\left(1+\frac{(\Delta-\Delta_1+\Delta_2)(\Delta+\Delta_3-\Delta_4)}{2\Delta}x+\dots\right)
\ee
where $\Delta=\frac{Q^2}{4}+P^2$.  These two terms represent the contributions to the correlation function from the exchange of the primary operator $V_{\alpha=\frac{Q}{2}+iP}$ and its $L_{-1}$ descendant, respectively.

The decomposition (\ref{eq:4ptLFTq}) can be interpreted as a representation of a sphere with four holes by gluing together two pairs-of-pants along a common cuff.  Each pair of pants is represented by a three point vertex $C(\alpha_1,\alpha_2, \alpha)$, and the integral over $P$ is a sum over possible lengths of the cuff which are glued together.  As we will review below, in the classical limit this is not just an analogy but an exact statement in hyperbolic geometry.

In addition to the correlation functions (\ref{corr}),  it is possible to study correlation functions of Liouville theory on a general Riemann surface $\Sigma$.  Even the vacuum amplitude---the partition function with no operator insertions---is non-trivial at higher genus, and has a  conformal block decomposition related in a similar manner to the pair-of-pants decomposition of $\Sigma$.

\subsection{Semiclassical Liouville theory and hyperbolic geometry}\label{s:semiclassics}

To make a connection to the classical geometry of constant negative curvature Riemann surfaces we will consider the semiclassical limit of the action (\ref{eq:SL}). This is the $c\to \infty$ (or $Q\to \infty$) limit. In this limit it is convenient to introduce a new field, $\tilde \phi := 2b\phi$ and the parameter $\tilde \mu:= \mu b^2$, so that the action of equation (\ref{eq:SL}) takes the form
\be\label{eq:SLc}
S_L[\tilde \phi]: ={1\over b^2} \int_\Sigma d^2z \sqrt g \left ({1\over 16\pi} g^{\alpha\beta} \partial_\alpha \tilde \phi\partial_\beta \tilde \phi + \tilde \mu e^{\tilde \phi}\right )~.
\ee
We will then take the semiclassical limit $b\to 0$ with $\tilde \mu$ is fixed. We see from (\ref{eq:SLc}) that the parameter $b$ plays the role of $\sqrt \hbar.$ With a flat reference metric $g$ (which we can enforce with an appropriate choice of boundary condition) the classical equation of motion for $\tilde \phi$ is Liouville's equation
\be\label{LiouvEqn}
\partial \bar \partial \tilde\phi=2\pi\tilde \mu e^{\tilde \phi}.
\ee

When we take this semiclassical limit we will also scale the dimensions of the Liouville vertex operators 
by taking $p=\f{P}{Q}$ fixed as $b\to0$.  
So the dimension will scale as
\be
\Delta_{\alpha} = \alpha(Q-\alpha) = \f{Q^2}{4} (1-\xi^2) \approx \f{1}{4b^2} (1-\xi^2),
\ee
in the semiclassical limit, where we have defined $\xi:=
i\f{Q}{2}P$. In particular,  we define the Liouville momentum $p$ in the semiclassical limit via the relation $\alpha = \f{Q}{2} + i P := \f{Q}{2}(1 + 2ip)$.
We also will sometimes refer to the conformal weight of a primary operator in the semiclassical limit by the rescaled dimension $\delta := {\Delta\over Q^2}= \f{1-\xi^2}{4}$, which is also used in \cite{Hadasz:2005gk}. Note that in the semiclassical limit, these correspond to heavy operators, as for fixed $\xi,\delta$, the original dimensions $\Delta$ scale like $Q^2$.

When we take the semiclassical limit in this way, the $n$-point correlation function is given by a classical action,
\be
\langle V_{\alpha_1}(z_1,\bar z_1) \ldots V_{\alpha_n}(z_n,\bar z_n) \rangle_{b\to 0}\to \exp\left ( -{1\over b^2} S_L^{cl}[\tilde \phi_*]\right),
\ee
where $S^{cl}_L[\tilde \phi_*]$ is the action (\ref{eq:SLclass}) evaluated at $\tilde \phi_*$, the unique solution of (\ref{LiouvEqn}) with certain prescribed boundary conditions on $\tilde \phi(z,\bar z)$ at the locations of the vertex operators $z_1,\ldots z_s$ on $\Sigma.$ (See, e.g., \cite{Teschner:2003at,Hadasz:2005gk} for the precise form of the boundary conditions.)

The connection to hyperbolic geometry then arises from the fact that the classical solution to (\ref{LiouvEqn}), $\tilde \phi_*(z,\bar z)$, furnishes a unique metric of constant negative curvature on the Riemann surface $\Sigma$ with $n$ elliptic or parabolic singularities (cone points or punctures, resp.) at the set of points $\{z_1, \ldots z_n\}$.  In particular, the metric,  
\be
ds^2 = e^{\tilde \phi_*(z,\bar z)}|dz|^2,
\ee
has constant negative curvature, because the Ricci scalar
\be
R = -4e^{-\tilde \phi}\partial  \bar{\partial} \tilde \phi ,
\ee
becomes constant $R= -8\pi \tilde \mu$ when the Liouville equation of motion is satisfied.  

Note that the type of singularity which appears at the point $z_i$ depends on the semiclassical conformal weight $\delta_i$ of $V_{\alpha_i}(z_i)$ in the following way: for $0<\xi_i<1$ ($0<\delta_i<1/4$), there is a conical singularity of opening angle $2\pi \xi_i$ at the point $z_i$.  Vertex operators where $\xi$ takes imaginary values correspond to holes. In terms of the hyperbolic metric, when $\xi_i = {i\ell_i\over 2\pi}$, corresponding to 
\be\label{eq:geolength}
\delta_i = {1\over 4} + {1\over 4} \left ( {\ell_i \over 2\pi}\right )^2,
\ee there is a hole with a geodesic boundary of length $\ell_i$.  The special case when $\xi_i\to 0$ ($\delta_i \to 1/4$) will be of particular interest for us, as in this case there is a puncture at the point $z_i$.

We will be interested in understanding the geometry of $\mathcal M_{0,4}$, the moduli space of four-punctured spheres, so we begin by considering
 the semiclassical limit of the four point function \eqref{eq:4ptLFTq}.  
 In order to compute the semiclassical limit of the four point function, we need to know both the semiclassical behavior of the three point function $C(\alpha_1,\alpha_2,\alpha_3)$, as well as the conformal block $\mathcal F_P$.

In the semi-classical limit the DOZZ formula for the structure constants reduces to an exponential of the classical Liouville action of the form  (see e.g. \cite{Zamolodchikov:1995aa,Harlow:2011ny})
\be
C(\alpha_1,\alpha_2,\alpha_3) \approx e^{-\f{1}{b^2} S^{cl}_3(\delta_1,\delta_2,\delta_3)},
\ee
where $ S^{cl}_3(\delta_1,\delta_2,\delta_3)$ is the classical 3-point action of vertex operators with dimensions $\delta_1, \delta_2,\delta_3$ on the sphere.
Furthermore, the conformal block was conjectured to take a similar exponential form in the semiclassical limit by Zamolodchikov \cite{Zamolodchikov:426555}, which has since been proven \cite{Besken:2019jyw},
\be
\mathcal{F}_P\begin{bmatrix}\Delta_3 & \Delta_2 \\ \Delta_4  & \Delta_1 \end{bmatrix} (x) \approx \exp \Bigg[\f{1}{b^2}\ f_{p} \begin{bmatrix} \delta_3 & \delta_2 \\ \delta_4 & \delta_1\end{bmatrix}(x)  \Bigg],
\ee
where $f_p$ is called the ``classical conformal block."  Although closed form expressions for $f_p$ are not known, it can be evaluated in an expansion at small $x$ simply by taking the large $c$ limit of the expansion in equation (\ref{eq:FP}).

Since both the structure constants and the conformal block take an exponential form, the semiclassical limit of the four point function can be written as,
\ba
&&\langle V_{\alpha_4}(\infty,\infty) V_{\alpha_3}(1,1)V_{\alpha_2}(x,\bar{x}) V_{\alpha_1}(0,0)   \rangle \notag \\
&& \approx \int dp \exp\Bigg[-\f{1}{b^2}  \Big(S^{cl}_3(\delta_1,\delta_2,  1/4 + p^2) + S^{cl}_3( 1/4 + p^2,\delta_3,\delta_4) - 2 \text{Re} f_p \begin{bmatrix} \delta_3 & \delta_2 \\ \delta_4 & \delta_1\end{bmatrix}(x) \Big)\Bigg], \notag \\
\ea
where $p$ is the Liouville momentum of the internal vertex operator.
In the semiclassical limit we may approximate this integral using the method of saddle points, with saddle point determined by the 
equation
\be\label{Spointeq}
\f{d}{dp}\bigg(S^{cl}_3(\delta_1,\delta_2,  1/4 + p^2) + S^{cl}_3( 1/4 + p^2,\delta_3,\delta_4) - 2 \text{Re} f_p \begin{bmatrix} \delta_3 & \delta_2 \\ \delta_4 & \delta_1\end{bmatrix}(x)\bigg)\bigg |_{p=p_s} = 0.
\ee
The solution $p= p_s(x,\bar{x})$ to this equation determines the saddle point value of the internal Liouville momentum in terms of the cross ratio. 
Finally, inserting this saddle point value into the action, we arrive at the semiclassical limit of the Liouville four point function,
\ba
&&\langle V_{\alpha_4}(\infty,\infty) V_{\alpha_3}(1,1)V_{\alpha_2}(x,\bar{x}) V_{\alpha_1}(0,0)   \rangle \approx e^{-\f{1}{b^2}S^{cl}_L(x,\bar{x})}, \label{eq:4ptsaddle} \\
&& S^{cl}_L(x,\bar{x}) := (S^{cl}_3(\delta_1,\delta_2,  1/4 + p_s^2) + S^{cl}_3( 1/4 + p_s^2,\delta_3,\delta_4) - 2 \text{Re} f_{p_s} \begin{bmatrix} \delta_3 & \delta_2 \\ \delta_4 & \delta_1\end{bmatrix}(x),\notag
\ea
which is a function of the cross ratio $x$.
In the context of the moduli space $\mathcal M_{0,4}$ this classical action has an important interpretation: it is the K\"ahler potential for the Weil-Petersson metric (see e.g. \cite{zograf1987liouville}).  We will discuss this in more detail below.

\begin{figure}[htb]
\begin{center}
\includegraphics[width=6cm]{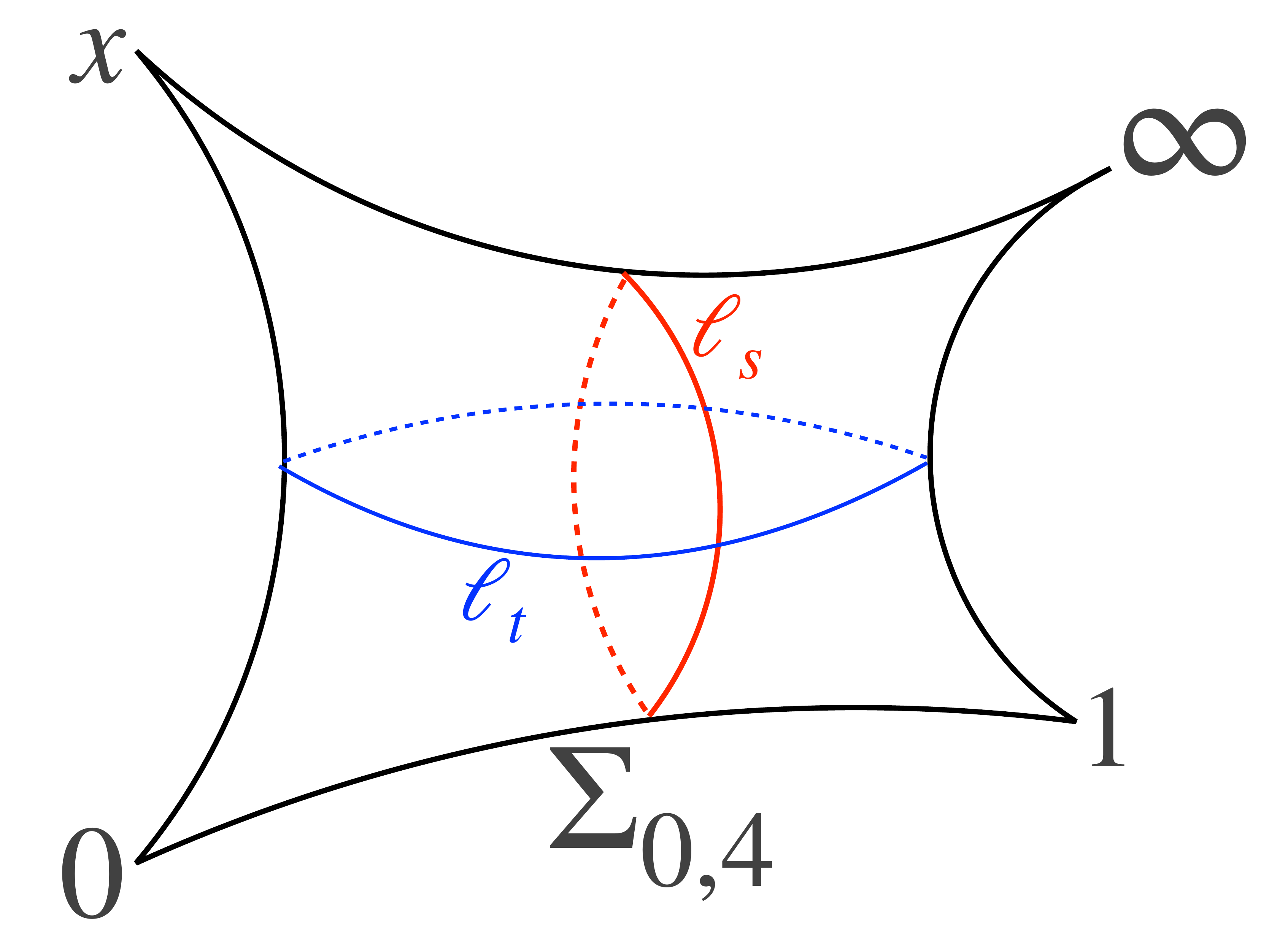}
\caption{Illustration of a four-punctured Riemann sphere $\Sigma_{0,4}$. 
In blue and red are the minimal $s$- and $t$-channel geodesics, of lengths $\ell_s$ and $\ell_t$, respectively. }\label{fig:sig04}
\end{center}
\end{figure}

We note that the saddle point momentum $p_s$ has the following physical interpretation---the action of equation (\ref{eq:4ptsaddle}) in the semiclassical limit describes a process where an operator of dimension $\delta_s= {1\over 4} + p_s^2$ is exchanged in the $s$-channel. From equation (\ref{eq:geolength}), we see that this corresponds to the length $\ell_s=4\pi p_s$  of the minimal $s$-channel geodesic on the four-punctured sphere, separating the two punctures (or conical singularities) at $0$ and $x$ from the two punctures located at $1$ and $\infty$. Similarly, a solution for the saddle point momentum in the $t$-channel or $u$-channel corresponds to lengths $\ell_t$ and $\ell_u$ of minimal geodesics around the punctures at $1$ and $x$ or $\infty$ and $x$, respectively. See Figure \ref{fig:sig04} for an illustration of the minimal $s$-channel and $t$-channel geodesics on the four-punctured sphere.

In \S \ref{sec:WPmetric} we will specialize to the case of a sphere with four punctures by taking $\delta_i={1\over 4}, ~ i=1,\ldots 4,$ and study the saddle point equation (\ref{Spointeq}). Using Zamolodchikov's recursion relations for the conformal blocks $f_p$, we will numerically approximate the saddle point action, which will allow us to numerically compute the Weil-Petersson metric on $\mathcal M_{0,4}$ to very high accuracy.

\section{The geometry of $\mathcal M_{0,4}$}\label{s:Modgeom}

Before proceeding, we will briefly summarize some aspects of the geometry of the moduli space $\mathcal M_{0,4}$ of the four-punctured sphere. 
\subsection{Hyperbolic Geometry}
We begin by recalling the construction of a Riemann surface with $n$ punctures $\Sigma_{0,n}$ as a quotient of the upper half plane. 
Let $\mathbb H$ be the upper half-plane with coordinates $\{(x,y)| x \in \R, y >0\}$ and hyperbolic metric,
\be \label{eq:Hmet}
ds^2 = {d x^2 + dy^2 \over y^2}.
\ee
The isometry group $PSL(2, \R)$ has a natural action on $\mathbb H$ given by,
$$ z \mapsto {az + b \over cz + d}, ~~ \begin{pmatrix} a & b\\c & d \end{pmatrix} \in PSL(2, \R),$$
where $z = x+ iy. $ Geodesics are circular arcs which intersect the real axis perpendicularly. The geodesic distance $d(z,w)$ between two points $z,w \in \mathbb H$ is fully determined by the trace of the matrix $g= \begin{pmatrix} a & b\\c & d \end{pmatrix}\in PSL(2,\R)$ which maps $z \mapsto w= {az + b \over cz + d}$ as,
\be\label{eq:leng}
\cosh {d(z,w)\over 2} = {1\over 2}\Tr g.
\ee

The uniformization theorem implies that a Riemann sphere with $n$ punctures $\Sigma_{0,n}$ is biholomorphic to the quotient $\mathbb H/G$ where $\mathbb H$ is endowed with the metric (\ref{eq:Hmet}) and $G< PSL(2,\R)$ is a Fuchsian (discrete) subgroup. The group $G$ is isomorphic to the fundamental group of $\Sigma_{0,n}$. Furthermore, it has $n$ generators $\gamma_i$, $i=1,\ldots n$, which satisfy $|\Tr \gamma_i|=2$, coming from the fact that each puncture corresponds to a length zero geodesic boundary of $\Sigma_{0,n}$.
Moreover, the generators satisfy the relation
\be\label{eq:monodromy}
\gamma_1\gamma_2\ldots \gamma_n= 1_{2\times 2},
\ee
due to the fact that any circle which encloses all $n$ punctures is contractible.

For any pair of punctures $z_i$ and $z_j$, there is a unique closed geodesic of the hyperbolic metric on $\Sigma_{0,n}$ which encloses precisely $z_i$ and $z_j$ but none of the other $n-2$ punctures. This geodesic has length $d_{ij}:=d(z_i,z_j)$ where $$\cosh {d_{ij}\over 2} = {1\over 2}|\Tr \gamma_i\gamma_j|.$$
Specializing to the case of four punctures, it is clear that a choice of $\Sigma_{0,4}$ depends on two real parameters\footnote{Each generator $a_i\in PSL(2,\R)$ contains two real degrees of freedom due to the constraint $|\Tr a_i|=2$. The additional constraint (\ref{eq:monodromy}) removes an additional 3 parameters. Finally, an overall conjugation symmetry by an element of $PSL(2,\R)$ removes an additional three degrees of freedom, such that we are left with $4\times 2-3-3=2$ real parameters.}, so the moduli space $\mathcal M_{0,4}$ is one-complex dimensional.

This moduli space enjoys several remarkable properties.  In particular, it is a complex manifold with a natural symplectic structure -- the Weil-Petersson symplectic structure -- with respect to which $\mathcal M_{0,4}$ is a K\"ahler manifold, with K\"ahler potential equal to the Liouville action $S^{cl}_L(x,\bar{x})$.  
  
There are several natural choices of coordinates for the moduli space $\mathcal M_{0,4}$, each of which are useful for illustrating different properties of the moduli space. For example, a simple set of coordinates are the lengths $\ell_s$ and $\ell_t$ of the minimal length geodesics in the $s$- and $t$-channel, as in Figure \ref{fig:sig04}.  If we think of the surface as the quotient $\mathbb H/G$, then these lengths are just given by formula (\ref{eq:leng}) for the corresponding elements of $G$.   One advantage of these coordinates these is that the Weil-Petersson symplectic structure can be explicitly computed, and takes the form
\be
\omega^{WP} = {1\over \cos \theta} d\ell_s \wedge d\ell_t,
\ee
where $\theta$ is the angle of intersection between the two geodesics.

Another related set of coordinates are the Fenchel-Nielsen coordinates, which are defined as follows.  
We start by noticing that there is a one-real parameter family of metrics on the disk with two punctures and a boundary that is a hyperbolic geodesic; this family of metrics is labelled by the hyperbolic length $\ell$ of the boundary circle.  We can then construct a four-punctured sphere by gluing  two of these disks together along their hyperbolic boundaries.  For example, we can think of the two disks as the left- and right-hand sides of Figure \ref{fig:sig04}, and the identify $\ell$ as the $s$-channel length $\ell_s$.  We are also free to glue together these disks with any relative angle, meaning that there is an additional twist parameter $\theta_s$.  The pair $(\ell_s,\theta_s)$ are the Fenchel-Nielsen coordinates.  In these coordinates the Weil-Petersson symplectic structure takes the very simple form,
\be
\omega^{WP}= d\ell_s \wedge d\theta_s.
\ee

One important point is that the coordinates described above are not unique.  For example, by construction the twist coordinate  is periodic: $\theta_s\sim \theta_s+\ell_s$.  A rotation of the angle $\theta_s$ by this amount is known as a Dehn twist.  Indeed, one could choose {\it any} closed geodesic on $\Sigma_{0,4}$ and consider the corresponding Dehn twist.  The coordinates  $(\ell_s,\theta_s)$ (or $(\ell_s,\ell_t)$) would transform in a complicated way under this Dehn twist, but these two different values of the coordinates would in fact describe the same surface $\Sigma_{0,4}$.  This gives the moduli space the structure of a quotient.  In particular, the group generated by the Dehn twists is the mapping class group, the group of diffeomorphisms modulo those which are continuously connected to the trivial diffeomorphism.  The moduli space $\mathcal M_{0,4}$ can be viewed as a quotient of Teichm\"uller space (the simply connected space parameterized by all real values of the Fenchel-Nielsen coordinates) by the mapping class group. Much of the difficultly in the study of moduli space comes from the complicated behaviour of this group action.

\subsection{Complex Geometry}

One advantage of working with the geodesic length coordinates is that the Weil-Petersson symplectic structure can be explicitly computed.
However, these coordinates obscure the fact that $\mathcal M_{0,4}$ is a complex manifold.  
To see this, another natural choice of coordinate on moduli space is the complex cross--ratio $x$. If $\{z_1, z_2, z_3, z_4\} \subset \C^*$ are the locations of four marked points on the Riemann sphere, there exists a unique $SL(2,\C)$ transformation which maps this set of points to $\{x,0,1,\infty\} \subset \C^*$, where
$$x = {(z_1-z_2)(z_3-z_4)\over (z_1-z_4)(z_3-z_2)}.$$
This $x$ is a complex coordinate on the moduli space ${\cal M}_{0,4}$. The moduli space then inherits the usual complex structure from $\C$.  

The Weil-Petersson metric is the K\"ahler metric associated with the sympectic form $\omega^{WP}$.  Since this symplectic form is easily computed in the length-twist coordinates, in principle one could compute the metric exactly provided one knew how to compute the length $\ell_s(x,{\bar x})$ as a function of the cross-ratio.  Unfortunately  $\ell_s(x,{\bar x})$ does not have a simple closed form, although one can study the problem perturbatively in an expansion near $x=0$.  We will use essentially this method in the next section, although we will find it more convenient to work directly with the Liouville action rather than the length coordinates. 

Note that our original moduli space is the space of labelled points, so naturally admits the action of the group $S_4$ which permutes the four points. All of the structures which we are considering are invariant under this action.  The cross ratio $x$ is invariant under the subgroup $\Z_2 \times \Z_2 \in S_4$ which contains the products of two elements of order two in $S_4$.  The remaining permutations, which can be taken to live in the quotient $S_3 = S_4 / \left(\Z_2 \times \Z_2\right)$, act on $x$ as the anharmonic transformations,
\beq \label{eq:S3}
S_3: x\to \left\{ x, {1-x},{1\over x}, {1\over 1-x}, {x \over x-1},{x-1 \over x}\right\} .
\eeq
These are generated by $x\to 1-x$ and $x\to 1/x$, which are the usual crossing transformations that interchange the $s$-channel with the $t-$ and $u-$channels, respectively.

The anharmonic group $S_3$ is order $6$, so this action divides the complex plane into six fundamental domains.  Thus, rather than taking $\M_{0,4}$ to be the full complex $x$ plane, we may take our moduli space to be a $6$-fold cover of a fundamental domain for this $S_3$ action.  We will take as our fundamental domain,
\be
\Delta = \{ {\rm Re}(x) <1/2,~ |x-1| <1\}  \subset \C.
\ee
This is a convenient choice for any computation which involves an approximation scheme which is valid near $x=0$ but is expected to break down near the other singular points at $x=1$ and $x=\infty$, such as an $s-$channel expansion.

\bigskip
\noindent \emph{Zamolodchikov's $q$ coordinate}
\medskip

As we will see when we return to the connection to Liouville theory, following \cite{zamolodchikov1987conformal}, it will be useful to define a new coordinate on moduli space, $q$, as
\be \label{eq:qcoord}
q:= \exp\left (-\pi {K(1-x)\over K(x)}\right ),\ee
where $x$ is the cross-ratio coordinate discussed above, and 
$$
K(x) := \int_0^1 {dt \over \sqrt{(1-t^2)(1-xt^2)}}
$$
is the complete elliptic integral of the first kind. For $x \in \C^* \backslash \{0,1,\infty\}$, it follows that $|q(x)|<1$. Furthermore, the map $x \mapsto q(x)$ furnishes a uniformization of the 3-punctured sphere by the unit disk $\mathbb D$ with a hyperbolic metric. Near the boundary of moduli space $x\to0$, we have 
$q \sim x/16$.

\begin{figure}[ht]
\begin{center}
\includegraphics[width=6.2cm]{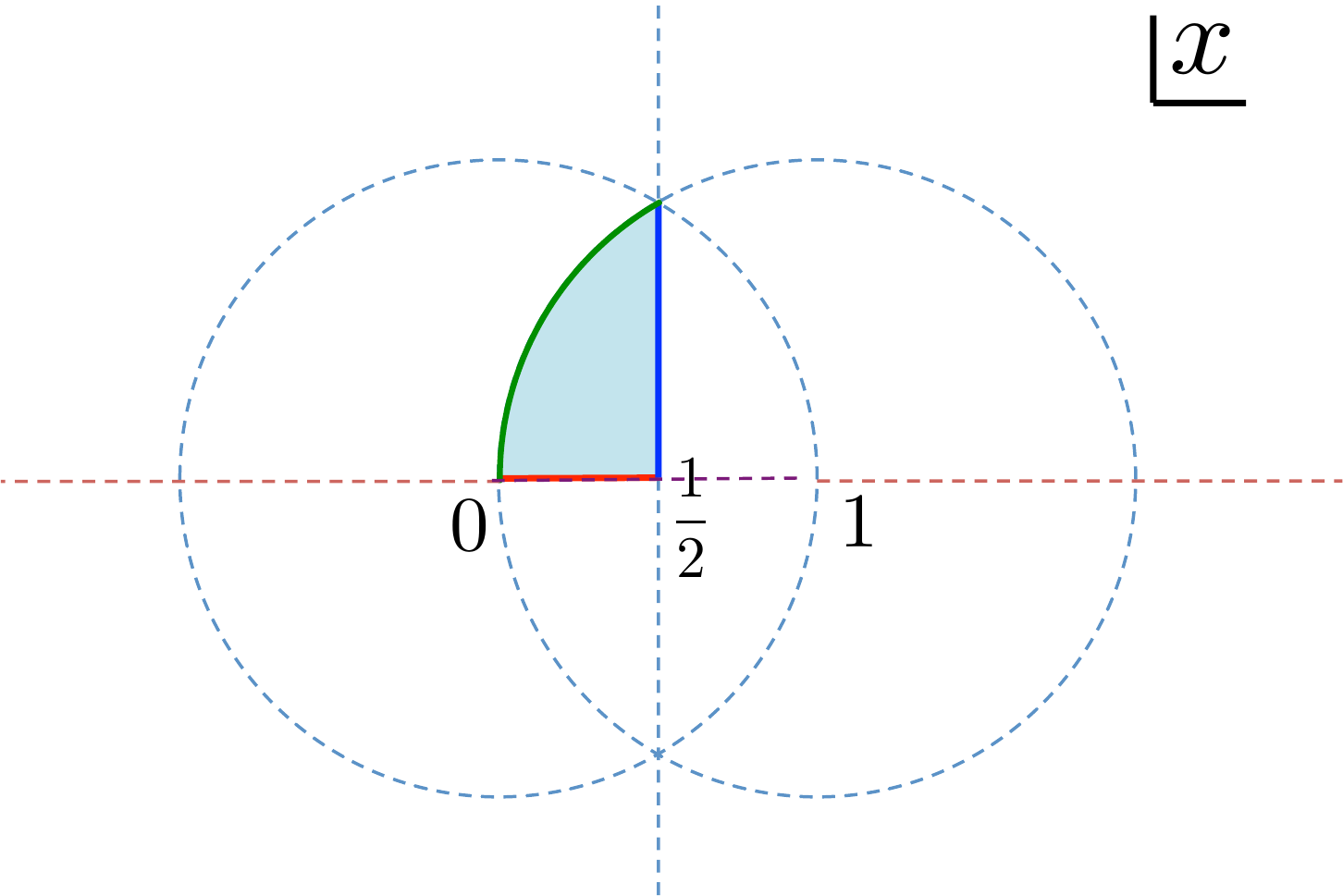}
\includegraphics[width=4.7cm]{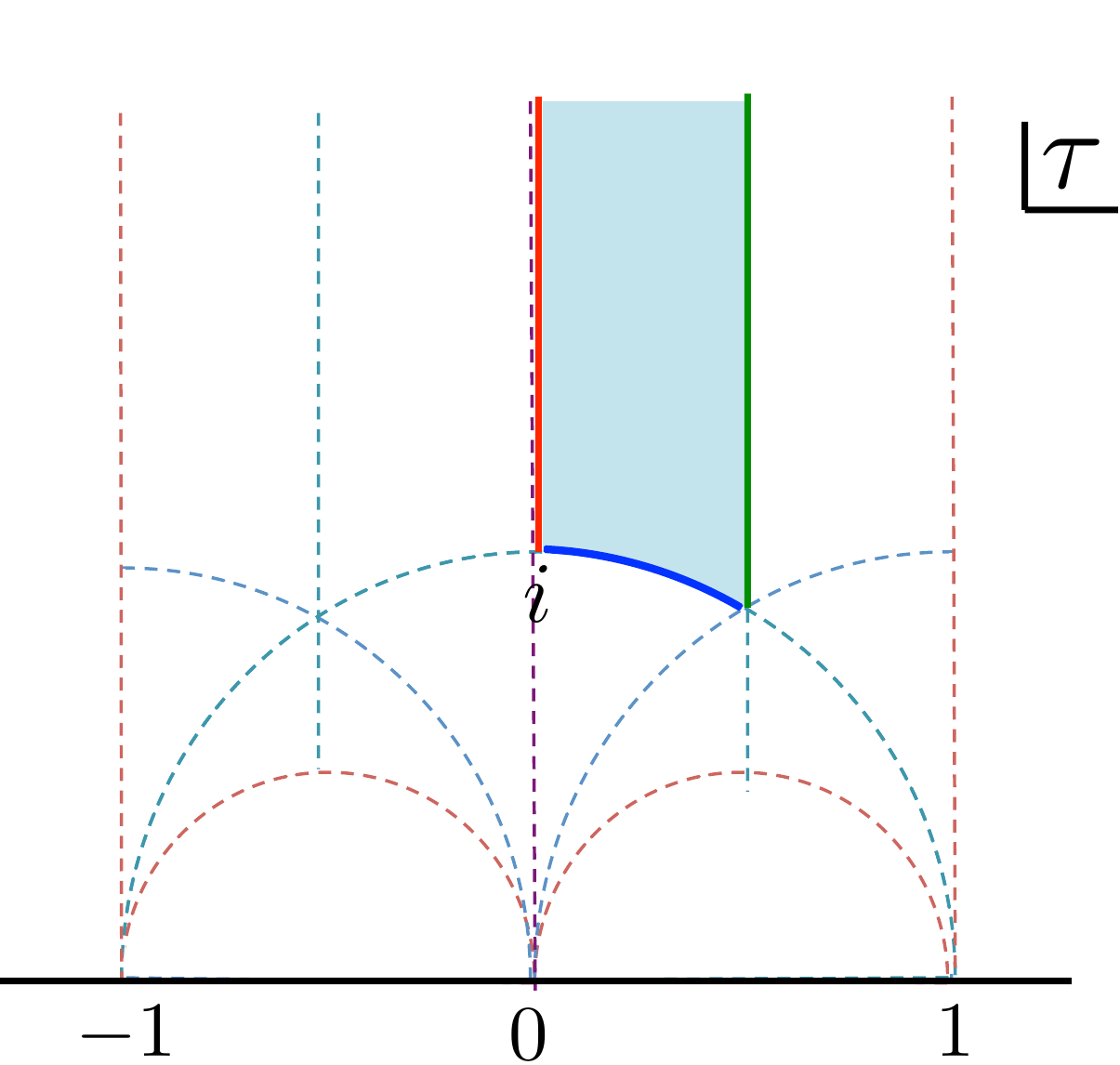}
\end{center}
\caption{The moduli space $\mathcal{M}_{0,4}$ in the cross-ratio coordinate $x$ (left) and the $\tau$ coordinate (right). 
The shaded region corresponds to half of the fundamental domain, and will later become the region on which we solve Laplace's equation. The full fundamental domain is found by considering the union of the red boundary line, the shaded region, and the reflection of the shaded region over the red line.}  
\label{fig:ModuliCoordinate}
\end{figure}

Finally, it will also be convenient to express $q$ as a function of a variable $\tau \in \mathbb H$, where $q = \exp(\pi i \tau)$, and consider $\tau$ as a complex coordinate on ${\cal M}_{0,4}$. It follows that in terms of the cross-ratio coordinate, $\tau = i{K(1-x) \over K(x)}$.
(Half-)fundamental domains for the $x$ and $\tau$ coordinates are pictured in Figure \ref{fig:ModuliCoordinate}.

\section{The Weil--Petersson metric on $\mathcal M_{0,4}$} \label{sec:WPmetric}

In this section, we will study the classical Liouville action $S_L^{cl}(x,\bar x)$ for a sphere with four punctures using the saddle point equation (\ref{Spointeq}). Since $S_L^{cl}(x,\bar x)$ is a K\"ahler potential\footnote{This was proven in  \cite{zograf1987liouville}, and also holds when the four singularities are elliptic (cone points) rather than parabolic (punctures).} for $\mathcal M_{0,4}$ this will allow us to study the the Weil--Petersson metric on the moduli space $\mathcal M_{0,4}$.  Our goal is to achieve an accurate numerical approximation for the metric.

Specifically, the 
Weil-Petersson metric on $\mathcal M_{0,4}$ is given by
\be
g_{x \bar{x}} = - 4 \pi \partial_{x}\partial_{\bar{x}}S_{L}^{cl}(x,\bar x),
\ee
where $S_L^{cl}(x,\bar x)$ is the classical Liouville action evaluated at the saddle point momentum $p_s(x)$. Once we have specialized to the case of four punctures, corresponding to $\delta_i={1\over 4}, ~ i=1,\ldots 4$, we can write the classical action as
\be\label{eq:SLclass}
S_{L}^{cl} (x,\bar x) = 2 S_3^{cl}(\tfrac{1}{4},\tfrac{1}{4},\tfrac{1}{4}+ p^2) - 2 \text{Re} \ f_{p} \begin{bmatrix}\f{1}{4} & \f{1}{4} \\ \f{1}{4} & \f{1}{4}\end{bmatrix}(x),
\ee
where $S_3^{cl}$ is the 3-point classical action and $f_p$ is the classical conformal block.

Furthermore, the 3-point classical action is explicitly given by \cite{Zamolodchikov:1995aa}
\be
S_3^{cl} (\tfrac{1}{4},\tfrac{1}{4},\tfrac{1}{4}+p^2) = \frac{1}{2} \log 2 + 2F (\tfrac{1}{2} + ip) +  2F (\tfrac{1}{2} - ip) - 2 F(0) + H (2ip) + \pi p,
\ee  
where we define 
\ba
F(x) &:=& \int _{\f{1}{2}} ^ x d y \log \f{\Gamma(y)}{\Gamma(1-y)} = \psi^{(-2)}(x) + \psi^{(-2)}(1-x) - 2\psi^{(-2)} (\tfrac{1}{2}), \notag \\
H(x) &:=& \int _0^x dy \log \f{\Gamma(-y)}{\Gamma(y)} = - \psi^{(-2)}(x) - \psi^{(-2)}(-x),
\ea
and $\psi^{(n)}(x)$ is the $n$-th polygamma function. Plugging this in to (\ref{Spointeq}), we see that the saddle point momentum $p_s(x)$ must satisfy the differential equation,

\ba
\frac{d}{dp} S_{L}^{cl} (x,\bar x;p)\Big |_{p=p_s} &=& \frac{d}{dp} \Big[ 2 S_3^{cl}(\tfrac{1}{4},\tfrac{1}{4},\tfrac{1}{4} +p^2) - 2 \text{Re} \ f_{p} \begin{bmatrix}\f{1}{4} & \f{1}{4} \\ \f{1}{4} & \f{1}{4}\end{bmatrix}(x) \Big]\Big |_{p=p_s}  \notag \\ \label{eq:p_s}
&=& - \pi +2 i \log \frac{\Gamma (1- 2ip_s) \Gamma^2(\tfrac{1}{2} + ip_s)}{\Gamma (1+ 2ip_s) \Gamma^2(\tfrac{1}{2} - ip_s)} - \text{Re} \f{\partial}{\partial p} f(p,x)\Big |_{p=p_s} = 0,
\ea
where we have defined the notation $f(p,x)$ by 
\begin{align}
f(p,x) &:= f_{p} \begin{bmatrix}\f{1}{4} & \f{1}{4} \\ \f{1}{4} & \f{1}{4}\end{bmatrix}(x) \notag \\
&= \Big(p^2 - \f{1}{4}\Big) \log x + \Big( \f{1}{8}  + \f{p^2}{2}\Big) x + \Big(\f{9}{128} + \f{13}{64}p^2 + \f{1}{1024 (1 + p^2)} \Big)x ^2 + O(x^3)
\end{align}
Here we have included the first few terms in the expansion of the classical conformal block.

In the rest of this section, we explicitly solve (\ref{eq:p_s}) for the saddle point momentum $p_s$ to increasing degrees of accuracy using Zamolodchikov's recursion for $f_p$, and use this to compute a numerical approximation to the Weil--Petersson metric $g_{x\bar x}$. In \S \ref{s:bdymetric}, we compute the metric near the boundary of moduli space, where the cross-ratio coordinate $x\to 0$. In \S \ref{s:gpert}, we derive what we term a ``perturbative" expansion of the saddle point momentum in terms of a variable $\vep(q, \bar q)\sim -1/\log(q\bar q)$, where $q$ is the coordinate defined in \S \ref{s:Modgeom}, which in principle allows us to compute the metric to arbitrarily high accuracy in the 3-point action, at leading order in the expansion of $f_p.$ We improve upon this in \S \ref{s:gnonpert}, where we outline a method to compute the saddle point momentum, and thus the metric, up to any desired order in the expansion of $f_p.$ As the expansion of $f_p$ amounts to an expansion of $\exp(-1/\vep(q,\bar q))$, we dub these ``non-perturbative" corrections. Finally, we also verify that our numerical approximation to the metric reproduces known quantities, such as the volume of $\mathcal M_{0,4}$ and geodesic lengths, to a very high degree of accuracy in \S \ref{s:vol}.

\subsection{Metric near the boundary of moduli space} \label{s:bdymetric}
We first analyze the behavior of the Weil-Petersson metric $g_{x\bar x}$ in the limit of small cross-ratio coordinate, $x\to 0$, which corresponds to a boundary of ${\cal M}_{0,4}$. This also corresponds to the limit where the geodesic length $\ell_s= 4\pi p_s \to 0$. 
In this limit the three-point function has the expansion, 
\be
S_3^{cl}\left({1\over 4}, {1\over 4}, {1\over 4} + p^2\right) \sim  \log 2 +  4 \psi^{(-2)}(1/2) -2\psi^{(-2)}(1) - \pi p + 8 p^2 \log 2  + \mathcal{O}(p^3).
\ee
This follows from the expansion of the functions,
\bea\nn
&F(1/2 + ip) \sim  -\psi(1/2) p^2 + \mathcal{O}(p^3),\\\nn
&H(2ip) \sim -2\pi p  - 4\gamma p^2 + \mathcal{O}(p^3),
\eea
near $p=0$, where $\psi(y)$ is the digamma function, which satisfies $\psi(1/2)+ \gamma=-2\log 2$, as well as the expansion of the conformal block near $p=0$,
\be
f(p,x) \sim 1 - \f{1}{4}\log x + p^2 \log x + \mathcal{O} (p^3). 
\ee
Therefore, at leading order in $p$, we can write the saddle point equation as,
\be
\f{\partial }{\partial p} (2 S_3 ^{cl}(\tfrac{1}{4},\tfrac{1}{4},\tfrac{1}{4} + p^2)  - f(p,x)- \bar{f}(p, \bar x))\Big|_{p=p_s} =- 2\pi +32 p_s\log 2 - 2 p_s\log x - 2p_s\log \bar{x} + \mathcal{O}(p_s^2)  = 0~.
\ee

The solution to this equation at leading order
\be
p_s(x,\bar x) = \f{\pi}{-\log (\bar{x}x) + 16 \log 2},
\ee
gives the value of the $s$-channel saddle point momentum at leading order near the boundary of moduli space, $x\to 0$. Note that when $|x| \ll 2^{-r}$, we can ignore the factor of $16\log 2$, and the length of the s-channel geodesic is approximately
\be
\ell_s = 4\pi p_s \approx \f{4\pi^2}{\log \f{2^{16}}{x \bar{x}}} \approx \f{2\pi^2 }{\log \f{1}{|x|}}.  
\ee

Plugging the saddle point momentum $p_s$ back into the expression (\ref{eq:SLclass}) for the Liouville action, we find 
\be
S_L^{cl}(x,\bar x)=-2 + 2 \log 2 -4 F(0) +\f{1}{4} \log x + \f{1}{4}\log \bar{x} +  \f{\pi^2 }{-\log (2^{-16}x \bar{x} )},
\ee
in the limit $x\to 0$.  This leads to the following leading order behavior of the Weil-Petersson metric,
\be
\boxed{g_{x\bar x}=-4\pi \partial _{x}\partial_{\bar{x}}S_{L}^{cl}(x,\bar{x}) = \f{8\pi^3}{x \bar{x} \log^3 (\f{2^{16}}{x \bar{x}} )}}
\ee
 near the boundary of moduli space, $x\to 0.$
This formula matches the expression for the Weil-Petersson metric near the boundary of moduli space derived by Wolpert \cite{1104159880} (section 2.3). 
Indeed, it is straightforward to check that this matches Wolpert's formula for the symplectic structure $\omega^{WP}= d\theta \wedge d\ell$ if one uses the fact (proven in \cite{1104159880}) that near  the boundary of moduli space
$$ \ell\sim \f{2\pi^2}{\log(1/|q|)}, ~~\f{2\pi \theta}{\ell}\sim {\rm arg}(q). $$

\subsection{Perturbative expansion of the metric }\label{s:gpert}
Our method will be to compute the metric in a perturbative expansion in the $s$-channel Liouville momentum around $p_s=0$. As we saw previously, the limit $p_s\to 0$ corresponds to the boundary of $\mathcal {M}_{0,4}$ in the cross-ratio coordinate, $x \to 0$.  This is a good approximation only for a small range of $x$, since the Taylor expansion for $f(p,x)$ converges only when $x <1$. 
So instead of using the cross-ratio coordinate we will study the saddle-point equation as a function of the variable $q(x)$, defined in \eqref{eq:qcoord}.  The advantage is that $f(p,x)$ will then rapidly converge at all points $x \in \mathbb C$, except for small neighborhoods around $x=1, \infty.$ 

\subsubsection*{The leading order metric}

We begin by writing the leading order expansion of the conformal block as,
\be
f(p,x) \sim -\f{1}{4} \log x - \f{1}{4}\log (1-x) -\f{7}{4} \log \f{2}{\pi} K(x) + p^2 \log 16 + p^2 \log q(x) + \mathcal{O}(p^3),
\ee
so that we can write the Liouville action up to $\mathcal{O}(p^2)$ as
\bea
S_L^{cl}(x,\bar x)&=& G(x) + \bar{G}(\bar{x}) + p^2 \log \f{2^8}{q(x) \bar{q}(\bar{x})} - 2 \pi p \notag \\\label{eq:SLOp2}
&=&  G(x) + \bar{G}(\bar{x})  + \log \f{2^8}{q(x) \bar{q}(\bar{x})} \Big(p - \f{\pi}{\log \f{2^8}{q(x) \bar{q}(\bar{x})}}\Big) ^2 - \f{\pi^2 }{\log \f{2^8}{q(x)\bar{q}(\bar{x})}},
\eea
where we have defined 
\be
G(x) := \log 2 + 4 \psi^{(-2)}(1/2) - 2\psi^{(-2)}(1) + \f{1}{4} \log x + \f{1}{4} \log (1-x),
\ee
which is a holomorphic function of $x$, and thus will not contribute in the computation of the metric. 
It is clear from the expression (\ref{eq:SLOp2}) that at leading order the saddle-point value of the momentum is, 
\be\label{eq:p(q)leading}
p_s (x) = \f{\pi }{\log \f{2^8}{q(x)\bar{q}(\bar{x})}},
\ee
which corresponds to the following length of the $s$-channel geodesic,
\be
l_s = 4\pi p_s \approx \f{4\pi^2}{\log \f{2^8}{q \bar{q}}} \underset{q\to 0}{\approx} \f{2\pi^2 }{\log \f{1}{|q|}}.  
\ee

Let $g_{q\bar q}$ be the Weil--Petersson metric in the $q$-coordinate, so that $ds^2= 2g_{q\bar q}dq d\bar q$. After evaluating the classical action at the value of $p_s$ at leading order in $q$ (\ref{eq:p(q)leading}), and taking two derivatives, we arrive at the following expression for the metric, 
\be \label{eq:leadg(q)}
\boxed{g_{q\bar q}=\f{8\pi^3 }{q \bar{q} (\log\f{2^8}{q\bar{q}})^3}.}
\ee
This is the result for $g_{q\bar q}$ at leading order in the $q$-expansion of $f(p,x)$. 
Note that (\ref{eq:leadg(q)}) can be expanded in terms of $-\log(q\bar q)$ as,
\be
g_{q\bar{q}} = \f{8\pi^3}{[-\log (q\bar{q})]^3} \Bigg[1 + \f{3 \log 2^8}{\log (q\bar{q})} + \cdots \Bigg]~.
\ee
This will contain infinitely many corrections in the $\f{1}{\log (q\bar{q})}$ expansion.

\subsubsection*{The metric at next-to-leading order} To find the metric at next-to-leading order, we begin by expanding $S_3^{cl}\left({1\over 4}, {1\over 4}, {1\over 4}+p^2\right)$  to $\mathcal{O}(p^4)$, so that the full action takes the form
\be
S_L^{cl}(q,\bar q;p)\sim -2\pi p + 8 p^2 \log 2 - 4\zeta (3) p^4  - p^2 \log (q\bar{q}) + G(q) + \bar G(\bar{q}) + \mathcal{O}(p^5),
\ee
where we have suppressed the dependence of $q$ on $x$, and, as previously, $ G(q)$ is a holomorphic function of $q$ and independent of $p$. At this order in $p$, the saddle point equation becomes,
\be
\f{\partial S_L^{cl}(q,\bar q;p)}{\partial p}\Big |_{p=p_s} \sim -2\pi + 32p_s \log 2 -16 \zeta (3) p_s^3 - 4p_s \log 16 - 2p_s\log q\bar{q} = 0,
\ee
which has the solution, 
\be
p_s \sim \f{\pi}{\log \f{2^8}{q\bar{q}}} + \f{8\zeta (3) \pi^3}{(\log \f{2^8}{q\bar{q}})^4}.
\ee

At this point, it will be convenient to define an expansion parameter $\vep(q,\bar{q}) := \f{\pi}{\log \f{2^8}{q\bar{q}}}$, which captures the leading order value of the saddle-point momentum, and such that $p_s$ is given by a power series in $\vep(q,\bar q)$ of the form,
\be \label{eq:pofep}
p_s \sim \vep(q,\bar{q}) + \f{8 \zeta(3)}{\pi} \vep(q,\bar{q})^4 + \cdots.
\ee
Plugging this value of $p_s$ back into the Liouville action, we find the on-shell value takes the form,
\be
S_L^{cl}(q,\bar{q};p_s(q)) = - \pi \vep(q,\bar{q}) - 4\zeta(3) \vep(q,\bar{q})^4+\cdots = - \f{\pi^2}{\log \f{2^8}{q\bar{q}}} - 4\zeta(3) \f{\pi^4}{(\log \f{2^8}{q\bar{q}})^4} + \cdots.
\ee
Taking two derivatives allows us to compute the Weil--Petersson metric at next-to-leading order in $\vep(q,\bar q)$, and we find,
\be
\boxed{g_{q\bar{q}} = -4\pi\partial_q \partial_{\bar{q}}S_L^{cl}(q,\bar{q}) 
= \f{8\pi^3}{q \bar{q} (\log \f{2^8}{q\bar{q}} )^3} \Bigg[ 1 +  \f{40 \pi^2 \zeta(3)}{ (\log \f{2^8}{q\bar{q}} )^3}+ \cdots \Bigg],}
\ee
which includes the first correction to the leading order result found in (\ref{eq:leadg(q)}).

\subsubsection*{A perturbative expansion for $g_{q\bar q}$}

Now we outline a general method for computing the metric $g_{q\bar q}$ to arbitrary order in $\vep(q,\bar q)$, at leading order in the expansion of $f(p,x)$. 
We begin by expanding the Liouville action as,
\be
S_L^{cl}(q,\bar{q};p) = -2\pi p + 8 p^2 \log 2   - p^2 \log (q\bar{q}) + 16 p^4\sum_{k=1}^{\infty} \f{(-1)^{k-1}(1-4^k)\zeta(2k+1)}{(2k+1)(2k+2)} p^{2k-2}  + G(q) + \bar{G}(\bar{q}) + \mathcal{O}(q),
\ee
where the coefficients in the sum over $k$ come from an explicit expansion of the three-point classical action $S_3^{cl}\left({1\over 4}, {1\over 4}, {1\over 4}+p^2\right)$.
Differentiating this action with respect to $p$ leads to the saddle point equation,
\ba
0&=&\f{\partial S_L^{cl}(q,\bar{q};p)}{\partial p} \\ \notag
 &=& -2\pi  + 16 p \log 2   - 2p \log (q\bar{q}) + 16p^3\sum_{k=1}^{\infty} \f{(-1)^{k-1}(1-4^k)\zeta(2k+1)}{(2k+1)} p^{2k-2}  + \mathcal{O}(q).
\ea
Rewriting this equation in terms of the previously defined parameter $\vep(q,\bar{q})$ leads to the relation,
\be
\vep(q,\bar{q}) = p + 	\f{8}{\pi} \vep(q,\bar{q}) p^3 \sum_{k=1}^{\infty} \f{(-1)^{k-1}(1-4^k)\zeta(2k+1)}{(2k+1)} p^{2k-2} + \mathcal{O}(q),
\ee
which we can once again invert to solve for $p_s$ as a power series in $\vep(q,\bar{q})$ of the form (\ref{eq:pofep}). With this value of the saddle point momentum in hand, we can then compute the on-shell action $ S_L^{cl}(q,\bar{q};p)$ and the metric $g_{q\bar q}$ to an arbitrarily high order in $\vep(q,\bar{q})$.

For example, the solution for the saddle point momentum up to  $\mathcal{O}(\vep^8)$ is
\ba
p_s(q,\bar{q}) &=& \vep(q,\bar{q}) + \f{8\zeta(3)}{\pi} \vep(q,\bar{q}) ^4 - \f{24\zeta(5)}{\pi} \vep(q,\bar{q}) ^6 + \f{192\zeta(3)^2}{\pi^2} \vep(q,\bar{q}) ^7 + \mathcal{O}(\vep^8). 
\ea
This leads to the following values of the Liouville on-shell action,
\be
S_L^{cl}(q,\bar{q};p) 
=  - \pi \vep(q,\bar{q}) - 4\zeta(3) \vep(q,\bar{q})^4 + 8\zeta(5)\vep(q,\bar{q})^6  -\f{64 \zeta(3)^2}{\pi}   \vep(q,\bar{q})^7 + \mathcal{O}(\vep^8), 
\ee
and the Weil-Peterssom metric,
\be\label{gqqp}
\boxed{g_{q\bar{q}} = \f{8\pi^3}{q \bar{q} (\log \f{2^8}{q\bar{q}} )^3} \Bigg[ 1 +  \f{40 \pi^2 \zeta(3)}{ (\log \f{2^8}{q\bar{q}} )^3}-  \f{168 \pi^4 \zeta(5)}{ (\log \f{2^8}{q\bar{q}} )^5} +\f{1792 \pi^4 \zeta(3)^2}{ (\log \f{2^8}{q\bar{q}} )^6} + \cdots\Bigg ],}
\ee
at this order.
Note that in this expansion the metric depends only on the norm $q\bar{q}$ and not on the phase of $q$.  This is an artifact of our expansion at the perturbative level, and will no longer hold once non-perturbative corrections are included.

\begin{figure}[ht]
\begin{center}
\includegraphics[width=7.8cm]{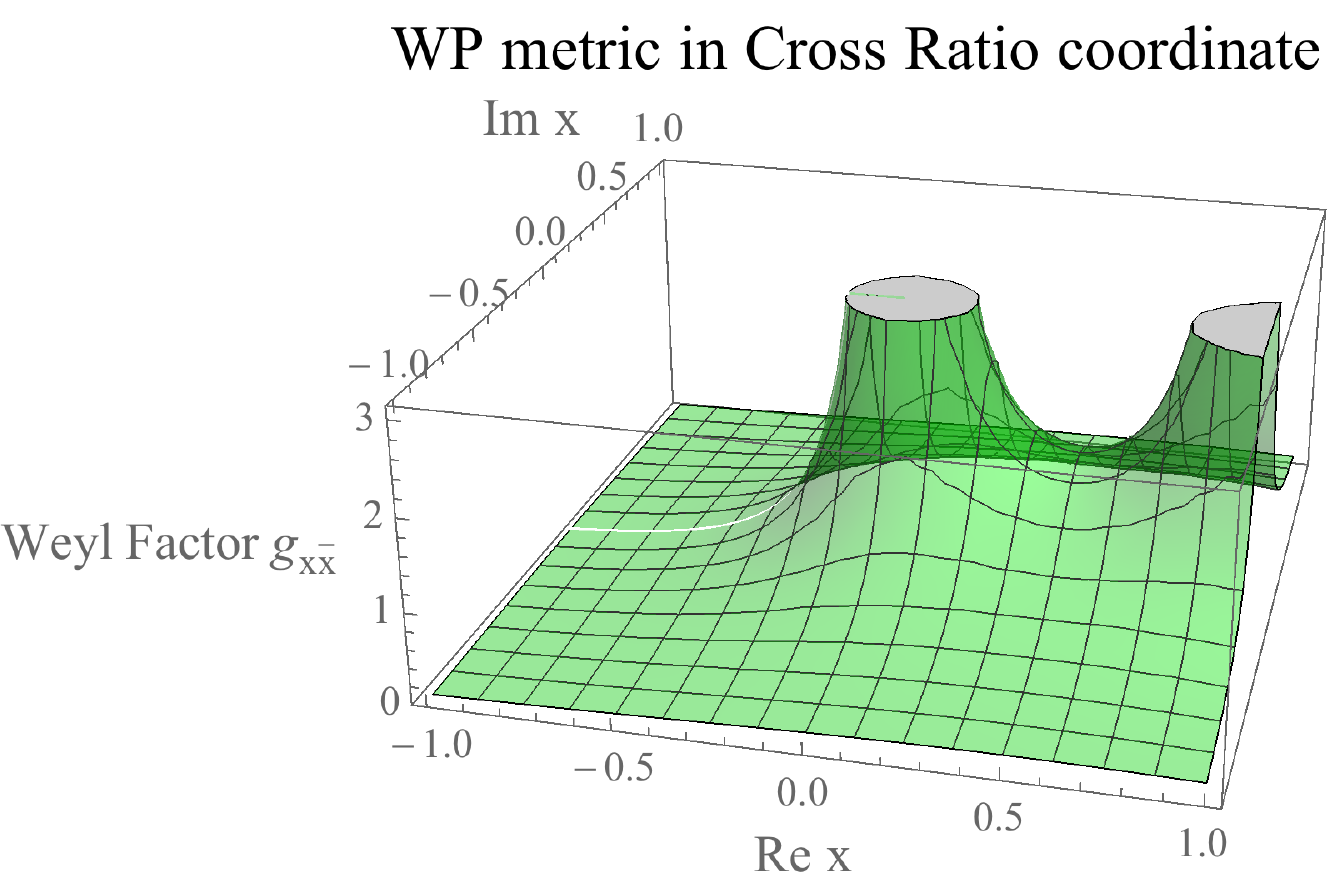}
\caption{A plot of the Weil--Petersson metric on $\mathcal{M}_{0,4}$. We include up to $O((\log(q))^{50})$ terms.}  
\label{fig:MetricPlotCRcoord}
\end{center}
\end{figure}

We plot the resulting expression for $g_{x\bar{x}}$ in the cross ratio coordinate in figure \ref{fig:MetricPlotCRcoord}.

\subsection{Non--perturbative corrections to the metric}\label{s:gnonpert}

Having discussed the solution to the saddle--point equation and the computation of the Weil--Petersson metric at leading order in the $q$-expansion of the conformal block $f(p,x)$ in the previous section, we now outline a method to incorporate higher-order terms $f(p,x)$ in these quantities. We refer to these terms as non--perturbative corrections since a power series in $q$ behaves like $\exp(-1/\vep(q,\bar q))$.

We begin by writing the $q$-expansion of the semiclassical conformal block as,
\ba
&&f_{\f{1}{4} + p^2}\begin{bmatrix}\f{1}{4} & \f{1}{4} \\ \f{1}{4} & \f{1}{4}\end{bmatrix}(x) \notag \\ 
&& \qquad = -\f{1}{4} \log x - \f{1}{4}\log (1-x) -\f{7}{4} \log \f{2}{\pi} K(x) + p^2 \log 16 + p^2 \log q(x)+ h_{\f{1}{4} + p^2}\begin{bmatrix}\f{1}{4} & \f{1}{4} \\ \f{1}{4} & \f{1}{4}\end{bmatrix} (q), \notag \\
\ea
where we introduce the function $h$ which includes the contribution to $f(p,x)$ which appears as a power series in $q$ (and which we dropped in our analysis in the previous section) and which has the expansion,
\ba
 h_{\f{1}{4} + p^2}\begin{bmatrix}\f{1}{4} & \f{1}{4} \\ \f{1}{4} & \f{1}{4}\end{bmatrix} (q)  
 &=& %1 +
  \f{1}{4(1+p^2)}q^2  \notag \\
 &&+ \Big( \f{-1}{32(1+p^2)^3} + \f{3}{128 (1+p^ 2) ^2 } + \f{15}{128(1+p^2)} + \f{81}{128(4+p^2)} \Big) q^4  \notag \\
 &&+ \Big( \f{1}{96(1+p^2)^5}- \f{5}{384(1+p^2)^4} -\f{9}{256 (1+p^2)^3} + \f{56}{2048(1+p^2)^2}\notag \\
 && \ + \f{661}{16384(1+p^2) }-\f{9}{128(4+p^2)} + \f{16875}{16384(9+p^2)}  \Big) q^6+ \cdots. 
 \notag \\
\ea
We denote this function $h(p,q)$ for simplicity. These formulas derive simply from the recursion relations of Zamolodchikov \cite{zamolodchikov1987conformal}.

Expanding $h(p,q)$ also as a power series in $p$, we can write
\ba
h(p,q) &=& %1 +
 \f{1}{4}q^2 + \f{137}{512} q^4 + \f{197}{1536} q^6 + \cdots  \notag \\
&& +  \Big( -\f{1}{4}q^2 - \f{225}{2048}q^4 - \f{5}{6144}q^6 + \cdots \Big)p^2 \notag \\
&& + p^4 \sum _{k=1} ^{\infty} A_k(q) p^{2k-2},
\ea
where $A_k(q)$ is a series of $q$ obtained from the $p$ expansion of the conformal blocks.
Note that every series is expanded as $A_k(q) = a_k q^2 + \cdots$.
The first few $A_k(q)$ are 
\ba
A_1(q) &=& \f{1}{4}q^2 + \f{81}{8192}q^4 - \f{12787}{221184}q^6 + \cdots, \notag \\
A_2(q) &=& -\f{1}{4}q^2 + \f{3247}{32768}q^4 + \f{732043}{7962624}q^6 + \cdots .
\ea
We also define the coefficients of the $p^0$ and $p^2$ terms in the expansion of $h(p,q)$ to be 
\be
A_{\text{indep}}(q) = %1 +
 \f{1}{4}q^2+ \f{137}{512} q^4 + \f{197}{1536} q^6 + \cdots 
\ee
and
\be
A_{\text{leading}}(q) = -\f{1}{4}q^2 - \f{225}{2048}q^4 - \f{5}{6144}q^6 + \cdots,
\ee
respectively.
Using this notation, the $p$-expansion of the conformal block is given by,
\be
h(p,q) = A_{\text{indep}}(q) + p^2 A_{\text{leading}}(q)  + p^4 \sum _{k=1} ^{\infty} A_k(q) p^{2k-2},
\ee
where each coefficient is also a power series in $q$ as described above.

Plugging in (\ref{eq:SLclass}), we can now write the full Liouville action as
\ba \label{eq:SLfull}
S_L^{cl}(q,\bar{q};p) 
&=&  G(q) + \bar{G}(\bar{q}) -2\pi p + 8 p^2 \log 2 - p^2 \log (q\bar{q}) -  (A_{\text{leading}}(q) + A_{\text{leading}}(\bar{q}))p^2 \notag \\
&&+ 16 p^4\sum_{k=1}^{\infty} \f{(-1)^{k-1}(1-4^k)\zeta(2k+1)}{(2k+1)(2k+2)} p^{2k-2}   - p^4 \sum _{k=1} ^{\infty} A_k(q) p^{2k-2} - p^4 \sum _{k=1} ^{\infty} A_k(\bar{q}) p^{2k-2}, \notag \\
\ea
where we have absorbed the $p$ independent piece $A_{\text{indep}}(q)$ into the pure  gauge  part $G(q)$. In the following we outline a procedure to solve the saddle-point equation and thus compute the metric after incorporating these additional corrections arising from the $p$-expansion of $h(p,q)$.

\subsubsection*{The expansion at leading order in $p$}
Note that now in (\ref{eq:SLfull}) the leading order piece in $p$ contains contributions from the non-perturbative corrections $A_{\text{leading}}(q)$.
The Liouville action up to order $p^2$ (and ignoring $p$ independent terms) becomes 
\ba
S_L^{cl}(q,\bar{q};p) 
&=& -2\pi p + p^2\Bigg[ \log \f{2^8}{q\bar{q}}-  (A_{\text{leading}}(q) + A_{\text{leading}}(\bar{q}))\Bigg] +\mathcal{O}(p^4),
\ea
leading to the saddle point equation at order $p$:
\be
\f{\partial }{\partial p}S_L^{cl}(q,\bar{q};p) = -2\pi p + 16 p \log 2 -  2p \log(q\bar{q}) -2p (A_{\text{leading}}(q) + A_{\text{leading}}(\bar{q})) = 0. 
\ee
This equation determines the saddle point value $p_s(q)$ as 
\ba
p_s(q)  &=& \f{\pi}{\log\f{2^8}{ q\bar{q}} -A_{\text{leading}}(q) - A_{\text{leading}}(\bar{q}) } \notag \\
 &=& \f{\pi}{ \log \f{2^8}{q\bar{q}} + \f{1}{4}( q^2 + \bar{q}^2) + \f{225}{2048} ( q^4 + \bar{q}^4) +  \f{5}{6144} (q^6 + \bar{q}^6)+ \cdots}.
\ea
Just as in the previous section, if we define a variable $\vep(q,\bar q)$ to be the leading-order (in $p$) value of the saddle point momentum, i.e.
\be
\vep(q,\bar{q}) := \f{\pi}{\log\f{2^8}{ q\bar{q}} -A_{\text{leading}}(q) - A_{\text{leading}}(\bar{q}) },
\ee
we can write the Liouville on-shell action at this order as 
\be
S_L^{cl}(q,\bar{q};p_s(q)) = -2 \pi p_s(q) + \pi\f{p_s(q)^2}{\vep(q,\bar{q})} = - \pi \vep(q,\bar{q}) 
\ee
Therefore, just by redefining $\vep(q,\bar{q})$ from its value in the previous section, we can represent both the saddle point equation and the Liouville action in the same manner.

Given this value of the classical action, we can compute the Weil-Petersson metric including corrections coming from the conformal block up to order $q^2$ (at this order $p^2$):
\be
\boxed{g_{q\bar q} 
= 4\pi \partial_q \partial_{\bar{q}}\Big( \f{\pi^2}{\log (\f{2^8}{q \bar{q}}) +\f{1}{4}q^2 + \f{1}{4}\bar{q}^2} + \cdots \Big)  dq d\bar{q}
= \f{128 \pi^3 (2 - q^2 )(2-\bar{q}^2)}{q\bar{q}(4 \log(\f{2^8}{q \bar{q}}) + q^2 + \bar{q}^2)^3} dq d\bar{q} + \cdots.}
\ee

\subsubsection*{Including subleading orders in $p$}

Finally, we briefly summarize the method to compute the metric including subleading terms in the expansion of $h(p,q)$.
We can express the full Liouville action (\ref{eq:SLfull}) (omitting the $p$ independent terms), including the newly-defined parameter $\vep(q,\bar q)$, as,
\ba
S_L^{cl}(q,\bar{q};p) &=& -2\pi p + \pi \f{p}{\vep(q,\bar{q})} \notag \\
&&+ 16 p^4\sum_{k=1}^{\infty} \f{(-1)^{k-1}(1-4^k)\zeta(2k+1)}{(2k+1)(2k+2)} p^{2k-2}   - p^4 \sum _{k=1} ^{\infty} A_k(q) p^{2k-2} - p^4 \sum _{k=1} ^{\infty} A_k(\bar{q}) p^{2k-2}, \notag \\
\ea
which leads to the full saddle point equation, 
\ba
\f{\partial}{\partial p}S_L^{cl}(q,\bar{q};p) &=& -2\pi + 2 \pi \f{p}{\vep(q,\bar{q})}  + 16 p^3\sum_{k=1}^{\infty} \f{(-1)^{k-1}(1-4^k)\zeta(2k+1)}{2k+1} p^{2k-2} \notag \\
&&- p^3 \sum _{k=1} ^{\infty} (2k+2) A_k(q) p^{2k-2} - p^3 \sum _{k=1} ^{\infty} (2k+2) A_k(\bar{q}) p^{2k-2} = 0, \label{eq:saddlenp1}
\ea
which we can again solve for $p$ recursively in $\vep$, by assuming a power series solution.

We forego explicitly solving this equation to any higher order than we have already discussed because---as we will see in a moment---the solutions for the metric at the orders we have  described already reproduce certain geometric quantities to extremely high accuracy.

Before doing so, we note that there is one final way that the accuracy of our approximation can be improved: we can use the $S^3$ symmetries of moduli space described in equation (\ref{eq:S3}).  These are isometries of the WP metric.
Our approximation for the metric is an expansion in cross-ratio, so will be most accurate when the cross ratio is small.  In particular it is most accurate in the fundamental domain depicted in figure \ref{fig:ModuliCoordinate}.  To obtain the metric at an arbitrary point in moduli space we will use an $S^3$ symmetry to map the point into the fundamental domain, and only then use our approximate form of the metric.  This is much more accurate than simply employing our approximation outside of the fundamental domain.

\subsection{Comparison: the Weil--Petersson volume of $\mathcal{M}_{0,4}$ }\label{s:vol}

We will now check that the approximations described above are accurate. 

We will begin by studying the volume of the moduli space $\text{Vol}(\mathcal{M}_{0,4})$ using our approximation for the metric $g_{q\bar q}$.
In the case of the moduli space of the four punctured sphere, the exact Weil--Petersson volume is \cite{wolpertmoduli},
\be
\text{Vol}(\mathcal{M}_{0,4}) = 2\pi^2.
\ee
Of course we can also compute the volume once we have $g_{q\bar q}$ on $\mathcal M_{0,4}$ by integrating the metric over the moduli space: 
\be
\text{Vol}(\mathcal{M}_{0,4}) = \int _{\mathcal{M}_{0,4}} g_{q\bar{q}}dq d\bar{q}.
\ee
We can therefore compare our expansion for $g_{q\bar q}$ from the previous sections to the exact result.

Furthermore, we can also test the accuracy of our result for the saddle point momentum $p_s(q)$ by computing the $s$-channel geodesic length at certain points in moduli space where the result is known exactly; in fact, this was already done in  \cite{Hadasz:2005gk}. 
Recall the relationship between the saddle point momentum and the geodesic length, $\ell_s= 4\pi p_s$, discussed in \S \ref{s:semiclassics}. At certain values of the cross ratio $x$, the quantity $\cosh\left ({\ell_s(x)\over 2}\right )$ is known exactly, and thus we can also compare $\cosh(2\pi p_s)$ using our approximation of $p_s$ to a given order.

\begin{table}[htb]
  \begin{center}
    \begin{tabular}{|c||c|c|c|c|c|} \hline
        & $\cosh (2\pi  p_s(1/2) )$  & $\cosh (2\pi  p_s(e^{i\f{\pi}{3}}) )$ & $g_{x\bar x}(1/2,1/2)$ & $g_{x\bar x}(e^{i\f{\pi}{3}},e^{-i\f{\pi}{3}})$ & Vol($\mathcal{M}_{0,4}$) \\   \hline
         $\vep$ &  2.74714 & $ 3.09768 $ & 1.23554 & 0.17457 & 17.5319 \\  \hline
      $\vep^4  $ & 3.00499 & 3.50136 & 1.58983 & 0.23704 & 19.701 \\ \hline
       $\vep^6  $ & 2.95594 & 3.41134 & 1.49928 & 0.218534 & 19.2835  \\ \hline
       $\vep^7  $ & 3.00215 & 3.5027 & 1.61307 & 0.243571  & 19.7552 \\ \hline
       $\vep^{18}  $ & 3.00082 & 3.50061 & 1.61307  & 0.243636   & 19.7455 \\ \hline
         $\vep^{60}  $ & 3.00039 & 3.49872 & 1.60833  & 0.242613  & 19.7374 \\ \hline
         Exact value  & $3 $  & $\f{7}{2}$ &  &  & $2\pi^2$  \\ \hline
    \end{tabular}
  \end{center}
  \caption{The value of $s$-channel geodesic length $\cosh (l_s/2) = \cosh (2\pi p_s(x))$ and the Weil-Petersson metric $g_{x\bar{x}}(x,\bar{x})$ in the cross ratio coordinate $x$ using the perturbative expansion of the saddle point momentum outlined in \S \ref{s:gpert}, and at leading order in the expansion of $f(p,x)$.
    $\vep^k$ is the order of the $\vep(q,\bar{q})$ expansion in $p_s$ and the Liouville action $S_L(p_s(q),q)$. The final column lists our results for the volume of $\mathcal M_{0,4}$.
    The exact volume is $2\pi^2 \approx 19.7392$.}\label{table:psmetric}
\end{table}

In Table \ref{table:psmetric}, we compare our approximation to the saddle point momentum $p_s$ using the methods of \S \ref{s:gpert} at two values of the cross ratio -- $x=1/2$ and $x=e^{i\f{\pi}{3}}$ -- to the exact values. We also compute the value of the metric $g_{x\bar x}$ in the cross ratio coordinate at these points, as well as our approximation of $\text{Vol}(\mathcal{M}_{0,4})$ using our result for the metric. We show results for different orders in the perturbative expansion in $\vep(q,\bar q)$ of $p_s$, at leading order in the conformal block expansion of $f(p,x)$. Once we include corrections up to $\mathcal O(\vep^{60})$, our approximate results differ from the exact values by about $10^{-3}$.

\begin{table}[htb]
  \begin{center}
    \begin{tabular}{|c||c|c|c|c|c|} \hline
        & $\cosh (2\pi  p_s(1/2) )$  & $\cosh (2\pi  p_s(e^{i\f{\pi}{3}}) )$ & $g_{x\bar x}(1/2,1/2)$ & $g_{x\bar x}(e^{i\f{\pi}{3}},e^{-i\f{\pi}{3}})$ & Vol($\mathcal{M}_{0,4}$) \\   \hline
         $\vep$ &  2.74714 & $ 3.09768 $ & 1.23554 & 0.17457 & 17.5319 \\  \hline
      $\vep^4  $ & 3.00458 & 3.50269 & 1.5868 & 0.238094 & 19.7028 \\ \hline
       $\vep^6  $ & 2.95555 & 3.41255 & 1.49643 & 0.219498 & 19.2851  \\ \hline
       $\vep^7  $ & 3.00174 & 3.50399 & 1.6101 & 0.244603  & 19.7570 \\ \hline
       $\vep^{18}  $ & 3.00042 & 3.501908 & 1.60752  & 0.244669   & 19.7445 \\ \hline
         $\vep^{60}  $ & 3.000000 & 3.500012 & 1.60537  & 0.243637   & 19.73922074 \\ \hline
         Exact value & $3 $  & $\f{7}{2}$ &  &  & $2\pi^2$  \\ \hline
    \end{tabular}
  \end{center}
  \caption{The value of $s$-channel geodesic length $\cosh (l_s/2) = \cosh (2\pi p_s(x))$ and the Weil-Petersson metric $g_{x\bar{x}}(x,\bar{x})$ in the cross ratio coordinate $x$ using the perturbative expansion of the saddle point momentum outlined in \S \ref{s:gnonpert}, including order $q^2$ non-perturbative corrections coming from the expansion of $f(p,x)$.
    $\vep^k$ is the order of the $\vep(q,\bar{q})$ expansion in $p_s$ and the Liouville action $S_L(p_s(q),q)$. The final column lists our results for the volume of $\mathcal M_{0,4}$.
    The exact volume is $2\pi^2 \approx 19.7392$. }\label{table:psmetricq2}
\end{table}

In Table \ref{table:psmetricq2}, we show the same results for geodesic lengths, the metric, and the volume of moduli space, now using the methods of \S \ref{s:gnonpert}. More specifically, we compute the solution of (\ref{eq:saddlenp1}) up to order  $(\vep(q,\bar{q}))^{60}$ including order $q^2$ non-perturbative corrections arising from the $q$-expansion of $f(p,x)$. We find that our results are extremely accurate even though we include just the first correction term in the expansion of the conformal block.
For example, we find a difference in the approximate volume of $\mathcal M_{0,4}$ and the true value at order  $(\vep(q,\bar{q}))^{60}$ of  
\be
\f{\text{Vol}(\mathcal{M}_{0,4})_{\text{approx}}}{2\pi^2} - 1 
\approx 6.04777\times 10^{-7}~.
\ee

\begin{figure}[ht]
\begin{center}
\includegraphics[width=9cm]{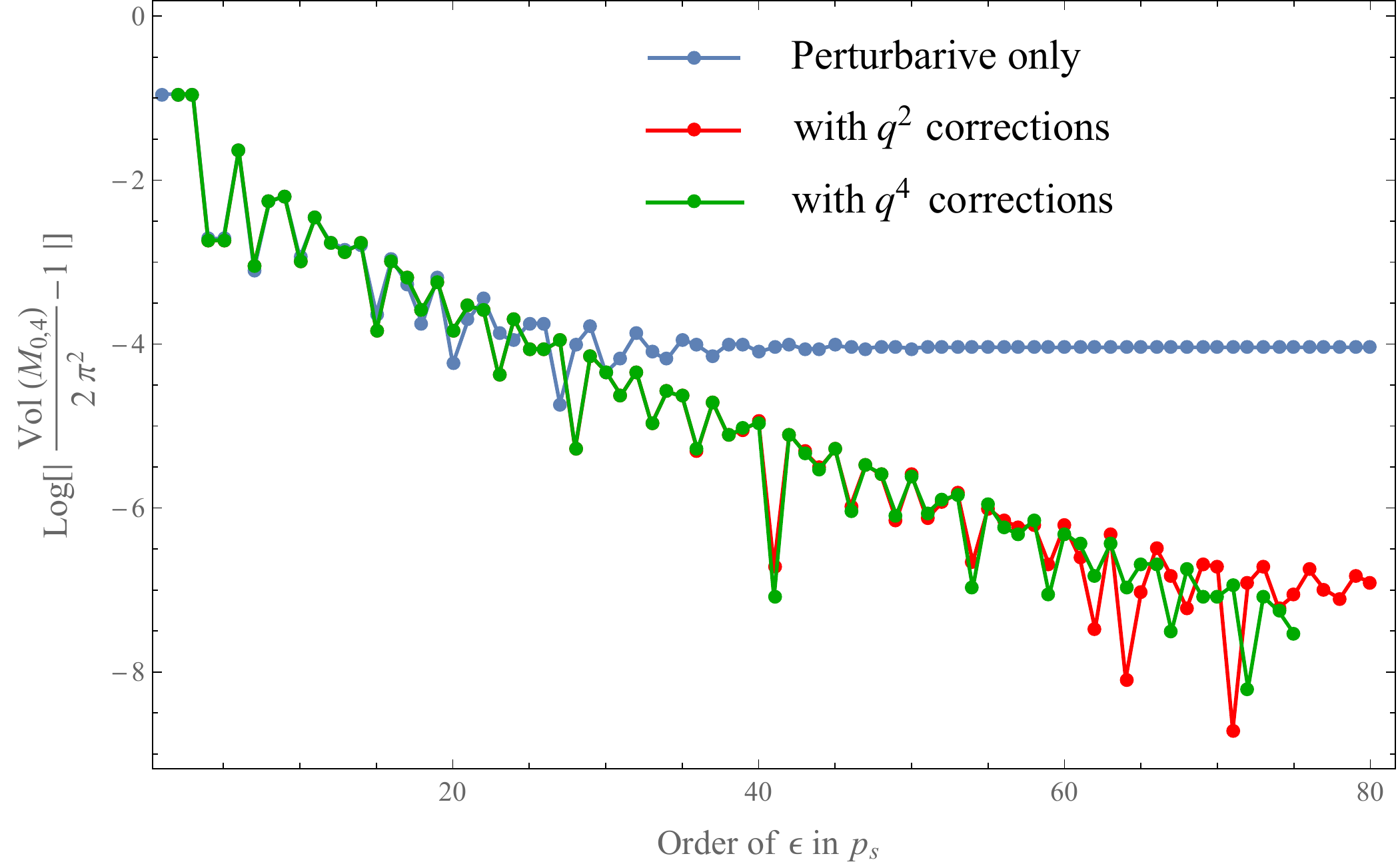}
\caption{Plot of $|\f{\text{Vol}(\mathcal{M}_{0,4})_{\text{approx}}}{2\pi^2} -1 |$ at each order of $\vep$.
Without non-perturbative corrections, the approximate volume converges around $19.7374$ and the difference is $|\f{\text{Vol}(\mathcal{M}_{0,4})_{\text{approx}}}{2\pi^2} -1 | \approx 9.1 \times 10^{-5}$.
Including non-perturbative corrections of order $q^2$, the difference continues to decrease after the perturbative results saturate.}  
\label{fig:Volumedif}
\end{center}
\end{figure}

Finally, in Figure \ref{fig:Volumedif} we plot the difference $|\f{\text{Vol}(\mathcal{M}_{0,4})_{\text{approx}}}{2\pi^2}-1|$ as a function of the order $\vep$, both with and without the including the $q^2$ term in the expansion of the conformal block $f(p,x)$. We see a clear decrease when we increase the order of $\vep$, which saturates at a particular value if we only include the perturbative corrections (leading order in $f(p,x)$), but continues to decrease substantially once we include the first nonzero term in the $q$-expansion of the conformal block.
We expect this pattern to continue, and that as more terms in the perturbative and non-perturbative expansions are included our results will continue to improve.

\section{The spectrum on moduli space}\label{s:spectrum}

In this section, we study the spectrum of the Weil--Petersson Laplacian on the moduli space $\mathcal{M}_{0,4}$. This describes the quantum mechanics of a particle moving on the moduli space $\mathcal M_{0,4}$ with Hamiltonian equal to the Weil--Petersson Laplacian, and can be viewed as a simple theory of quantum geometry/quantum gravity. We thus study the equation
\be\label{eq:eigen}
\Delta^{WP} \psi_n= E_n \psi_n,
\ee
where $\Delta^{WP}$ is the Laplacian constructed from the WP metric and $\psi_n$ is the $n$th eigenfunction, with corresponding eigenvalue $E_n.$

In the rest of this section, we present our numerical results on the values and statistics of the  $E_n$. 
We begin by briefly describing our method for numerically solving equation (\ref{eq:eigen}) in \S \ref{s:method}. 
For simplicity we will focus on the eigenfunctions which obey Dirichlet boundary conditions on the (half) fundamental domain depicted in Figure \ref{fig:ModuliCoordinate}. 
This is equivalent to restricting to the sector which is invariant under the $S_3$ symmetry of $\mathcal M_{0,4}$ and anti-symmetric under complex conjugation in $x$.\footnote{The methods used in this section can easily be adapted to study more general classes of eigenfunctions as well.} 
We will plot sample eigenfunctions $\psi_n$ on the (half) fundamental domain. In appendix \ref{app:eigenvalues}, we present explicit numerical values computed using the method described in \S \ref{s:method} in Tables \ref{table:SpectrumP65uv10} and \ref{table:SpectrumNP30uv100}.  Then we move on to analyze the statistics of the spectrum. In \S \ref{s:Weyl} we verify that our results obey Weyl's law, which describes the asymptotic eigenvalue density at large $n$. We next analyze the level statistics of the eigenvalues, computing the adjacent gap ratio (\S \ref{s:wigner}) and spectral form factor (\S \ref{s:SFF}).  These are probes of the spacings of eigenvalues, and in both cases we find behavior characteristic of the Gaussian Orthogonal Ensemble (GOE) universality class. This is evidence that the quantum mechanical system described above has quantum chaotic behavior.

\subsection{The spectrum of eigenvalues}\label{s:method}

\begin{figure}[ht]
\begin{center}
\includegraphics[width=5cm]{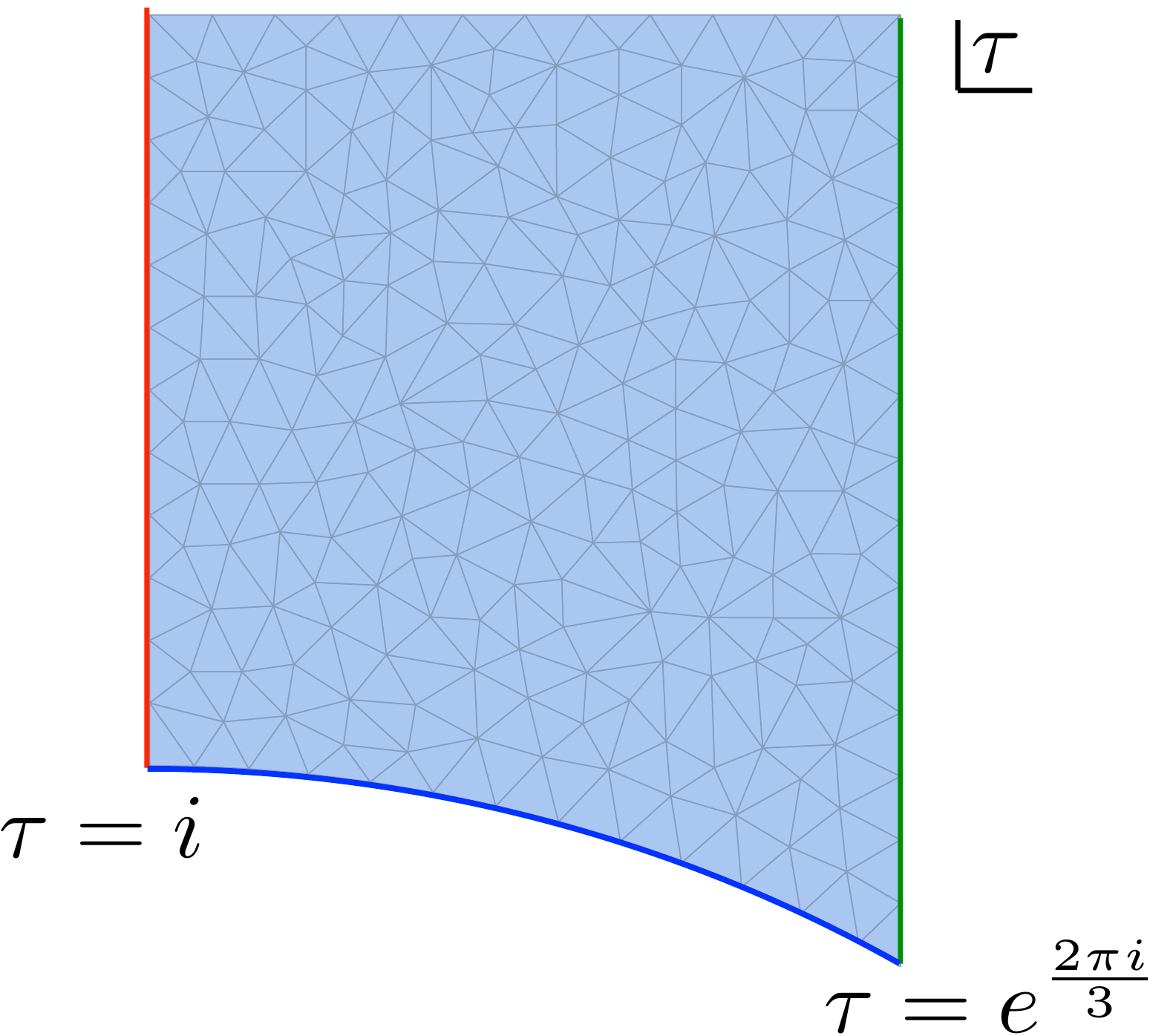}
\caption{An example of the discretization of half-fundamental domain of the moduli $\mathcal{M}_{0,4}$.
In this example, the area (with respect to the flat metric, not the WP metric) of each triangle is less than $0.001$.}  
\label{fig:DiscreteModuliM04}
\end{center}
\end{figure}

Using the expansion in $\vep\sim \log q$ and $q$ we developed in \S \ref{sec:WPmetric}, we can calculate the expansion of the metric analytically up to order $\mathcal O(q^m,(\log q)^n)$ for any desired values of $m$ and $n$.
We then proceed numerically by first discretizing the moduli space (with respect to the flat metric) such that the area of each element is less than some fixed value $A_{UV}$, and then numerically solving Laplace's equation. 
A pictorial example of such a discretization is illustrated in Figure \ref{fig:DiscreteModuliM04}, though the actual sizes of
$A_{UV}$ used in our calculations are $10^{-4}$ or $10^{-5}$, much smaller than that displayed in Figure \ref{fig:DiscreteModuliM04}. 

We work on the half-fundamental domain in the $\tau$ coordinate, introduced in \S \ref{s:Modgeom}. We first compute the WP metric to very high accuracy using the rapidly convergent $q$-expansion of the conformal blocks, and then convert the result to a function of $\tau$. In the $q$ variable, there is a high concentration of volume of the moduli space near the cusp $q=0$ (or correspondingly $x=0$ in the cross-ratio coordinate), and this leads to higher errors in the numerical solution of (\ref{eq:eigen}). By numerically solving (\ref{eq:eigen}) after converting to the $\tau$ coordinate, where this cusp is mapped to $\tau = i\infty$, this problem is avoided and we are able to achieve stable results for the eigenvalues at lower orders in $q$. 

As our goal is to analyze the statistics of the eigenvalue spectrum (rather than study the eigenvalues themselves), we find it convenient to make use of the following symmetries of the Laplacian.
Firstly, the Weil--Petersson metric in the cross-ratio coordinate is invariant under the $S_3$ symmetry described in (\ref{eq:S3}).
Additionally, there is a $\mathbb Z_2$ symmetry under complex conjugation, which exchanges $x \leftrightarrow \bar{x}$.  This is equivalent to $\tau \leftrightarrow - \bar{\tau}$.
The $S_3$ and complex conjugation symmetries commute with one another, and also commute with the Laplacian.
Thus when we solve Laplace's equation we can restrict to the sector where the wavefunction is symmetric under  $S_3$ and  anti-symmetric under the complex conjugation $\mathbb{Z}_2 := \{1,K\}$; i.e. the wavefunctions in this sector satisfy
\be
g \cdot \psi (x) = \psi(g^{-1}\cdot x) = \psi(x), \qquad \text{for } g \in S_3, \label{eq:S3sym}
\ee
and 
\be
K \cdot \psi (x) = \psi(K\cdot x) = -\psi(x), \qquad \text{for } K \in \mathbb{Z}_2. \label{eq:conjugtion}
\ee
This is equivalent to imposing Dirichlet boundary conditions on the boundary of the half--fundamental domain. 

\begin{figure}[ht]
\begin{center}\begin{tabular}{ccc}
\includegraphics[width=5cm]{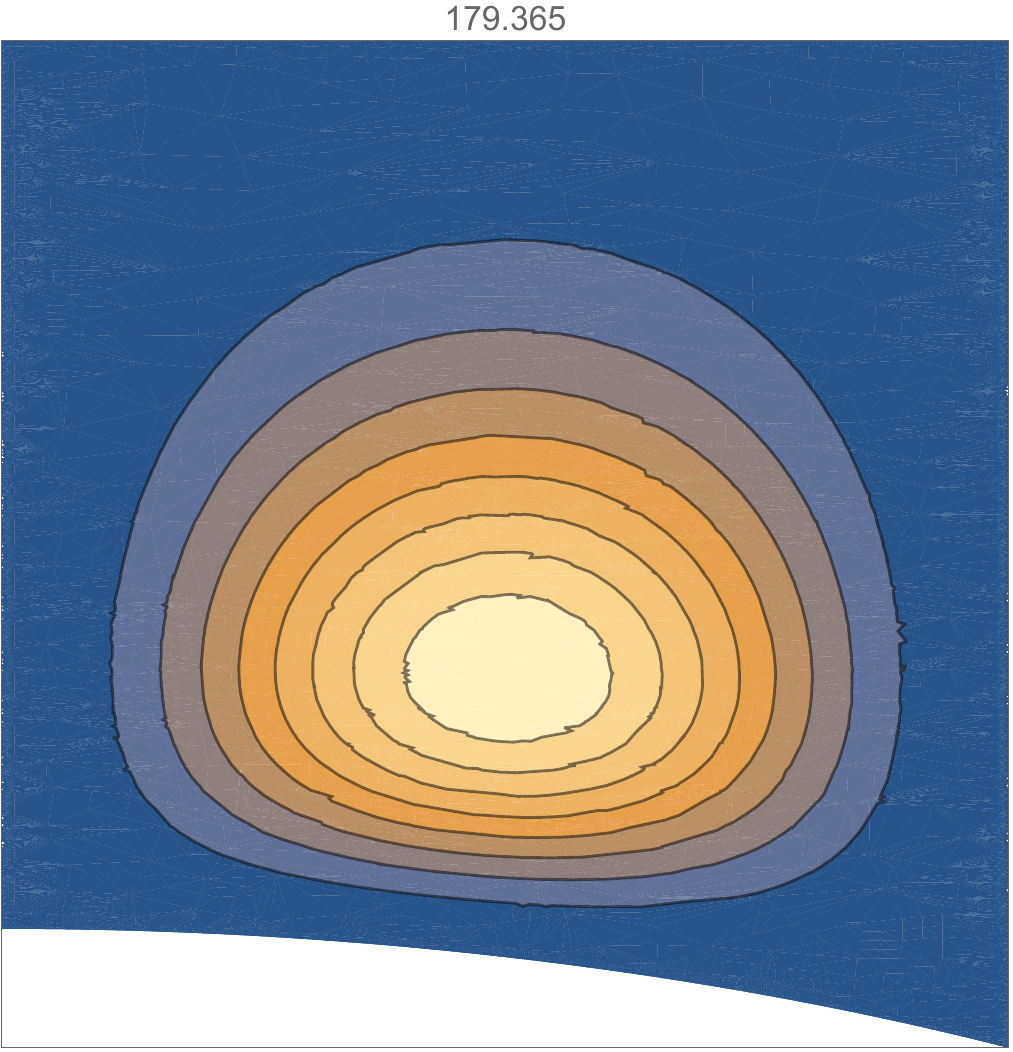}&
\includegraphics[width=5cm]{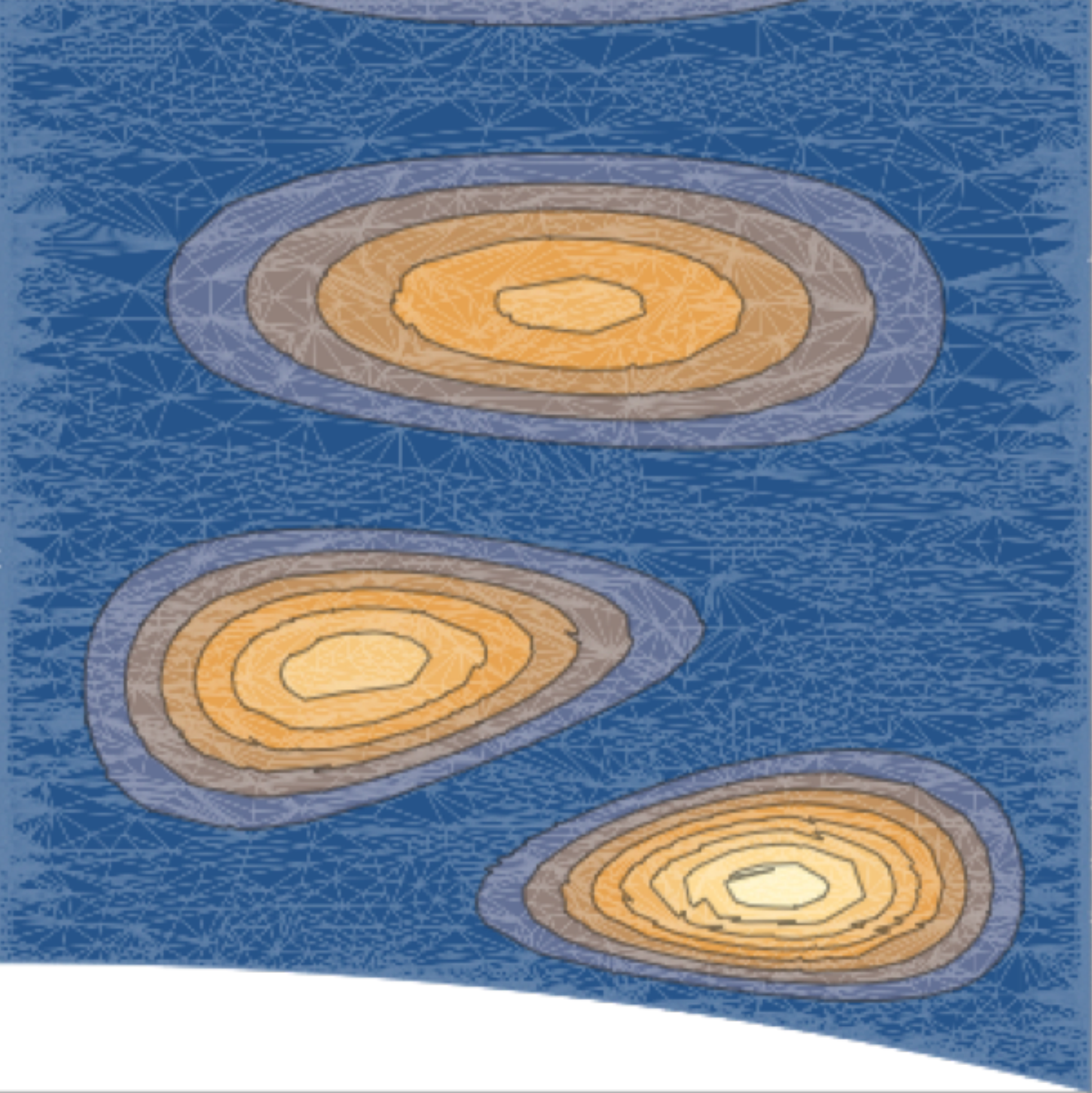}&
\includegraphics[width=5cm]{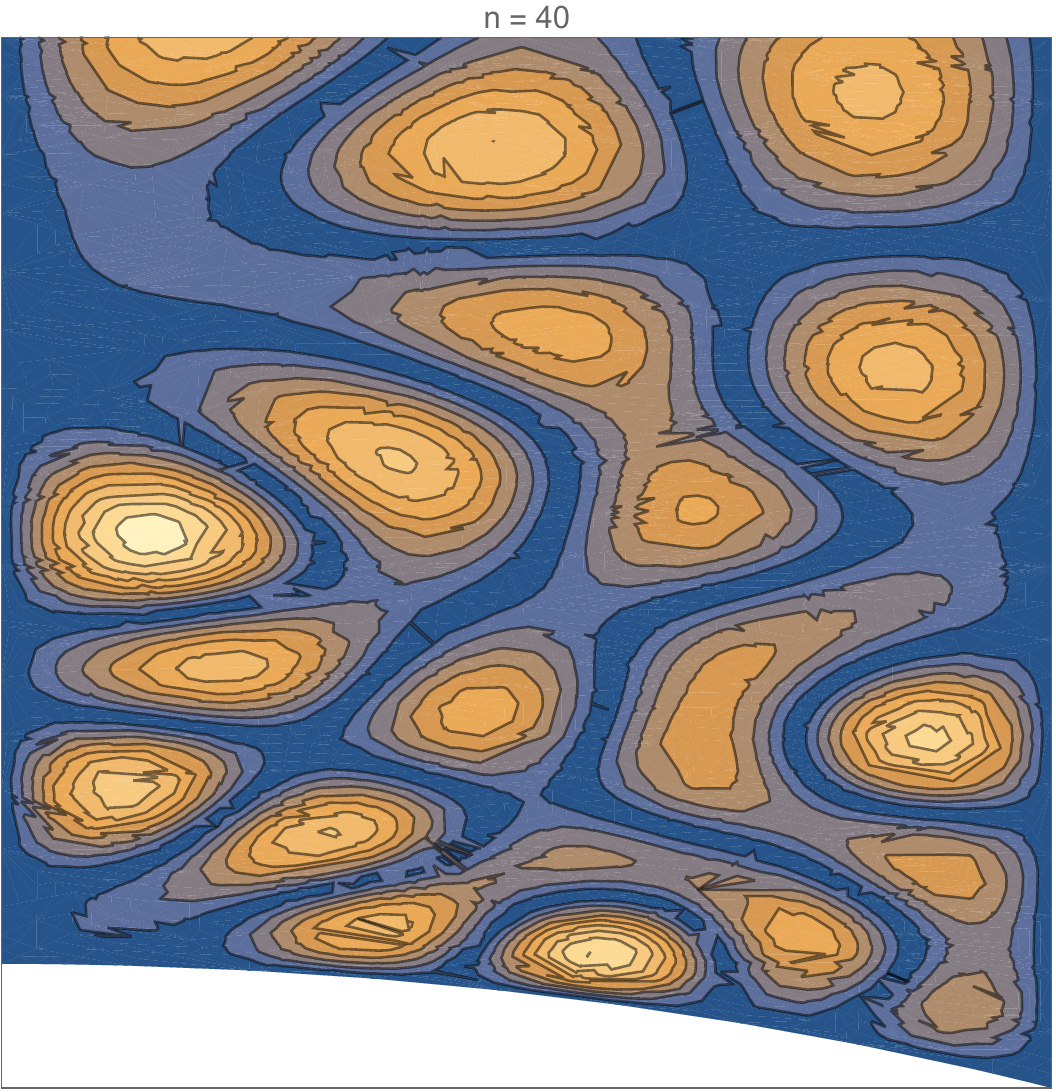}\\
$|\psi_1(\tau)|$& $|\psi_5(\tau)|$&$|\psi_{40}(\tau)|$ \end{tabular}
\caption{A contour plot of the absolute value of  our numerical results for selected eigenfunctions of the Laplacian $\Delta^{WP}$ on the half fundamental domain in the $\tau$ coordinate. On the  left is the ground state, $\psi_1(\tau)$, in the middle is the fifth excited state, $\psi_5(\tau)$, and on the right is the 40th excited state, $\psi_{40}(\tau)$. Note: these plots are cut off at a finite value of ${\rm Im}(\tau)$.}  
\label{fig:eigenfns}
\end{center}
\end{figure}

When we numerically solve (\ref{eq:eigen}), we obtain both the eigenvalues and the corresponding eigenfunctions of $\Delta^{WP}$. Though we are mainly interested in analyzing the statistics of the eigenvalues in the following sections, it is also interesting to observe the form of the eigenfunctions we obtain. We plot a few examples in Figure \ref{fig:eigenfns}.

\subsection{Asymptotic density of the spectrum: Weyl's law}\label{s:Weyl}
As a consistency check of our computation of the eigenvalues, we first verify  Weyl's law, which describes the asymptotic density of the eigenvalues as a function of energy.
More precisely, if we let $N(E)$ be the number of eigenvalues of the Laplacian of energy $\leq E$, then Weyl's law in two dimensions is given by (see, e.g., Ch. 16 of \cite{Gutzwiller})
\be
N(E) \approx \f{A}{4\pi} E \mp \f{L}{4\pi} \s{E} + K, \label{eq:WeylsLaw} 
\ee
where $A$ is the area of the manifold on which we are computing the Laplacian, $L$ is the length of the boundary and $K$ is a constant which depends on the topology of the manifold, the shape of its boundary, and the boundary conditions one implements.
Furthermore, the sign in front of the $\s{E}$ term depends on the boundary conditions: 
one has a positive or negative sign for Neumann or Dirichlet boundary conditions, respectively.

\begin{figure}[ht]
\begin{center}
\includegraphics[width=7cm]{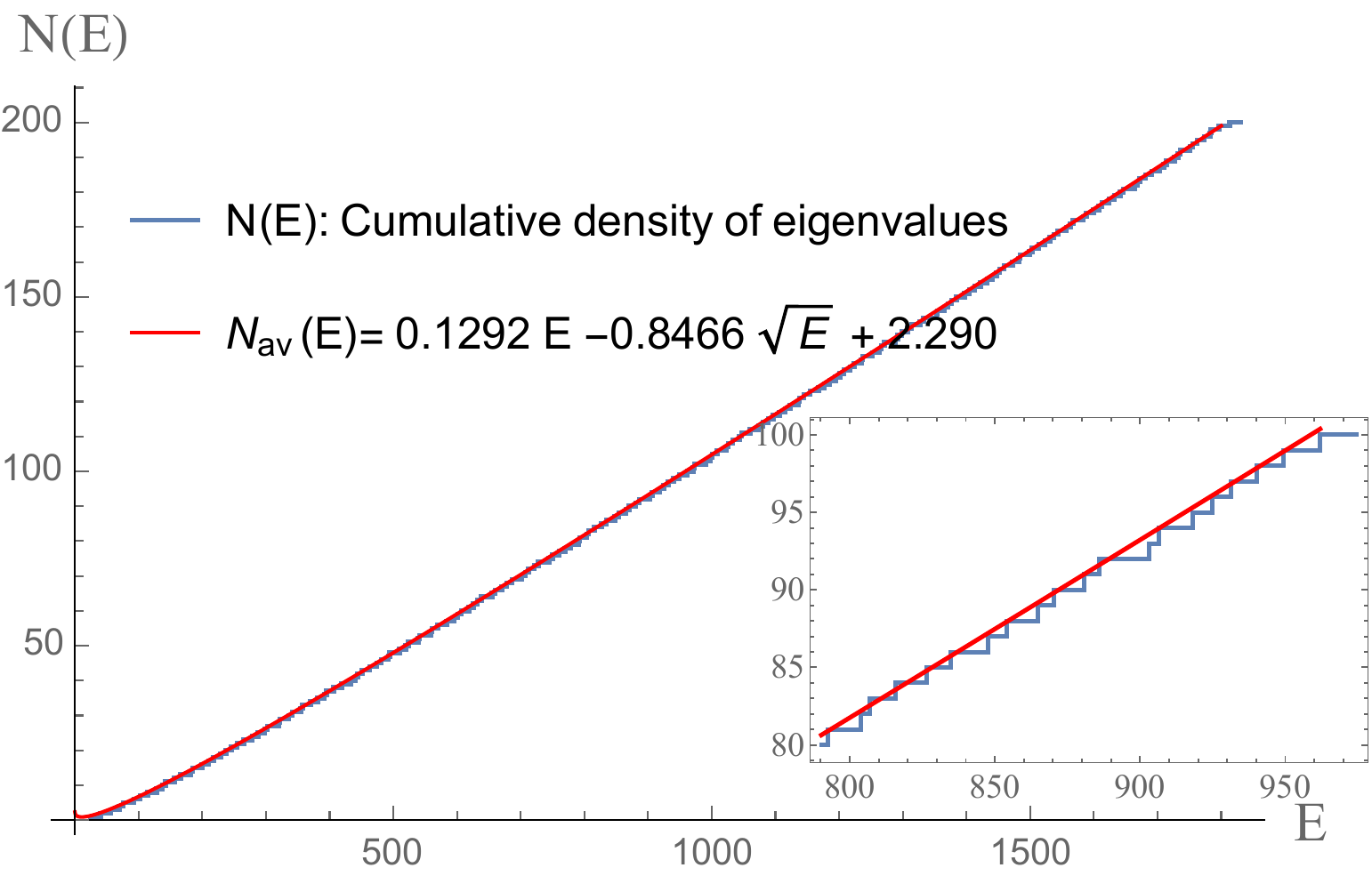}
\caption{The plot of the number of eigenvalues $N(E)$ with eigenvalue $\leq E$.
The blue line is the number of eigenvalues obtained numerically.
The red line is the fit with the function \eqref{eq:WeylsLaw}. Weyl's law gives the first coefficient as $\pi/24\approx 0.13.$
To compute eigenvalues numerically, we truncate the expansion at $q^0$ (without non perturbative corrections) and $(\log q)^{-60}$. The inset is an expansion of the plot between 80th to 100th eigenvalues. } 
\label{fig:WeylsLawPlot}
\end{center}
\end{figure}

In Figure \ref{fig:WeylsLawPlot}, we plot $N(E)$ for the first 200 numerically computed eigenvalues as a function of energy $E$, as well as the best fit of equation (\ref{eq:WeylsLaw}) to this distribution. Note that as we are solving Laplace's equation on a half-fundamental domain for the 6-fold symmetry group of moduli space, the expected coefficient of the linear term in $E$ in Weyl's Law (\ref{eq:WeylsLaw}) is 
$${A\over 4\pi}={1\over 4\pi} \times {1\over 2}\times {2\pi^2\over 6}={\pi \over 24} \approx 0.13~.$$
The best fit of the curve plotted in Figure \ref{fig:WeylsLawPlot} has coefficient $0.1292$, indicating strong agreement between our numerical results and Weyl's law.

\subsection{Level statistics: Nearest neighbor correlations}\label{s:wigner}

In this section we begin our study of the level statistics of the spectrum. We start by analyzing the short-range correlations in the spectrum of eigenvalues, by studying the distribution of nearest neighbor level spacings.

To set the stage, we recall that for $N\times N$ Gaussian orthogonal matrices, the joint probability distribution for the eigenvalues $E_i$, $i = 1,\dots, N$, is given by \footnote{{For more complete reviews, see \cite{mehta1991random, Guhr:1997ve} for example.}}
\be
P_N(E_1,\cdots,E_N) = C_N e^{-\f{\beta}{4} \sum_{i=1}^N V(E_i)} \Big( \prod_{1\le i,j \le N}|E_i - E_j|^{\beta} \Big),
\ee
with $V(E) = E^2$ and $\beta = 1$.
The spectrum of the WP Laplacian will not exactly match this joint probability distribution.  In particular, the quadratic potential $V(E)$ is specific to the Gaussian matrix ensemble.  For a general chaotic system the form of the potential might be quite different.
On the other hand, the Vandermonde determinant factor $\prod_{1\le i,j \le N}|E_i - E_j|^{\beta}$ depends only on the symmetry class ($\beta = 1 $ for GOE, $\beta = 2$ for GUE and $\beta =4$ for GSE) of the problem, and not on the detailed form of the matrix potential.  It describes a universal set of correlations between eigenvalues (essentially, eigenvalue repulsion) that is characteristic of any quantum chaotic system within that symmetry class.  We will study this universal behaviour in the spectral statistics of the WP Laplacian.

Instead of working directly with the distribution of nearest neighbor spacings between eigenvalues, $s_n: = E_{n+1}-E_n$, we work with the so--called adjacent gap ratio, $r_n$, defined by \cite{PhysRevB.95.115150,2013PhRvL.110h4101A},\footnote{Usually when one works directly with distribution of the nearest neighbor spacings $s_n$, one has to perform a process called ``unfolding" the spectrum. In performing the unfolding, one essentially divides by the local density of the spectrum in order to achieve a mean level spacing of unity which can then be compared to results from RMT.
A convenient feature of the adjacent gap ratio  \eqref{eq:defgapratio}--as described in \cite{2013PhRvL.110h4101A}--is that we do not need to perform this unfolding of the spectrum, because the average density dependence cancels after taking the ratio of the level spacings.}
\be
r_n := \f{E_{n+1}-E_n}{E_n-E_{n-1}}, \label{eq:defgapratio}
\ee
where $E_n$ is the $n$th energy eigenvalue of the system. Assuming our eigenvalues are drawn from an ensemble in RMT, the average of the adjacent gap ratio takes the form,
\be
\braket{r} = \int _0 ^{\infty} r p(r) dr ,
\ee
where $p(r)$ is the probability distribution of adjacent gap ratios. This quantity can be thought of as an analogue of Wigner's surmise for the probability distribution of nearest neighbor spacings, $p(s)$, for spectra derived from random matrix ensembles \cite{mehta1991random}. Results for the probability densities $p(r)$ and average adjacent gap ratios $\braket r$ can be found in \cite{2013PhRvL.110h4101A}, and we reproduce them here in Table  \ref{table:gapratio}.  The three RMT ensembles should be compared to Poisson statistics, the expected statistics of an integrable (non-chaotic) system.  The smaller value of $\braket r$ with Poisson statistics reflects the absence of eigenvalue repulsion in an integrable system.

\begin{table}[htb]
  \begin{center}
   
    \begin{tabular}{|c||c|c|c|c|} \hline
        &  Poisson  & GOE &GUE & GSE\\  \hline
         $p(r)$ &  $\f{1}{(1+r)^2}$ & $ \f{27}{8} \f{(r+r^2)}{(1+r+r^2)^{\f{5}{2}}} $ &$\f{81\s{3}}{4\pi} \f{(r+r^2)^2}{(1+r+r^2)^{4}}$ & $\f{729}{4\pi} \f{(r+r^2)^4}{(1+r+r^2)^{7}}$ \\  \hline
      $\braket{r}  $ & $2\log 2-1 \approx 0.38629$ & $4 - 2\s{3} \approx 0.53590$ &$2\f{\s{3}}{\pi} - \f{1}{2} \approx 0.60266$ & $\f{32}{15}\f{\s{3}}{\pi} - \f{1}{2} \approx 0.67617$  \\ \hline
    \end{tabular} \caption{Probability distribution $p(r)$ and average $\braket r$ of the adjacent gap ratio in systems whose eigenvalues obey either Poisson statistics, or statistics of one of the three ensembles from RMT \cite{2013PhRvL.110h4101A}. }\label{table:gapratio}
  \end{center}
\end{table}

\begin{figure}[ht]
\begin{minipage}{0.49\hsize}
\begin{center}
\includegraphics[width=7cm]{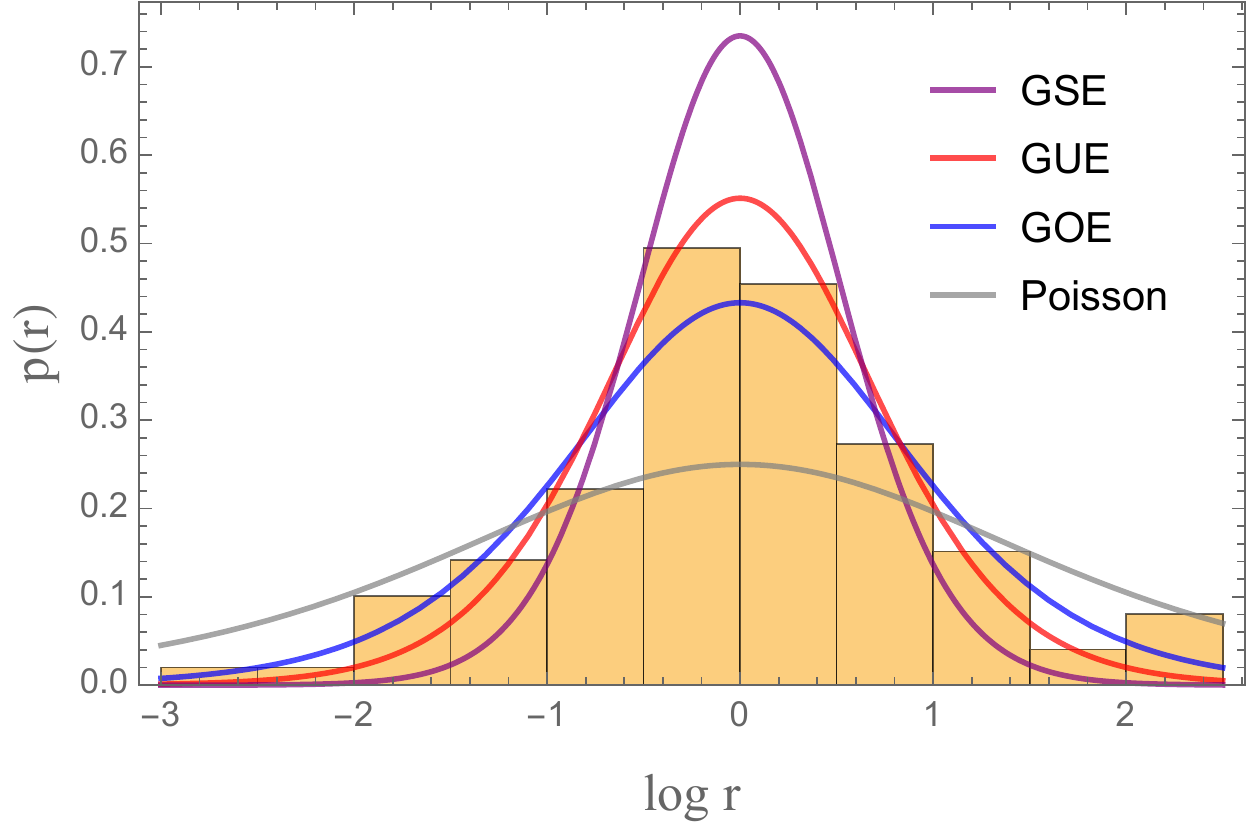}
\end{center}
\end{minipage}
\begin{minipage}{0.49\hsize}
\begin{center}
\includegraphics[width=6.2cm]{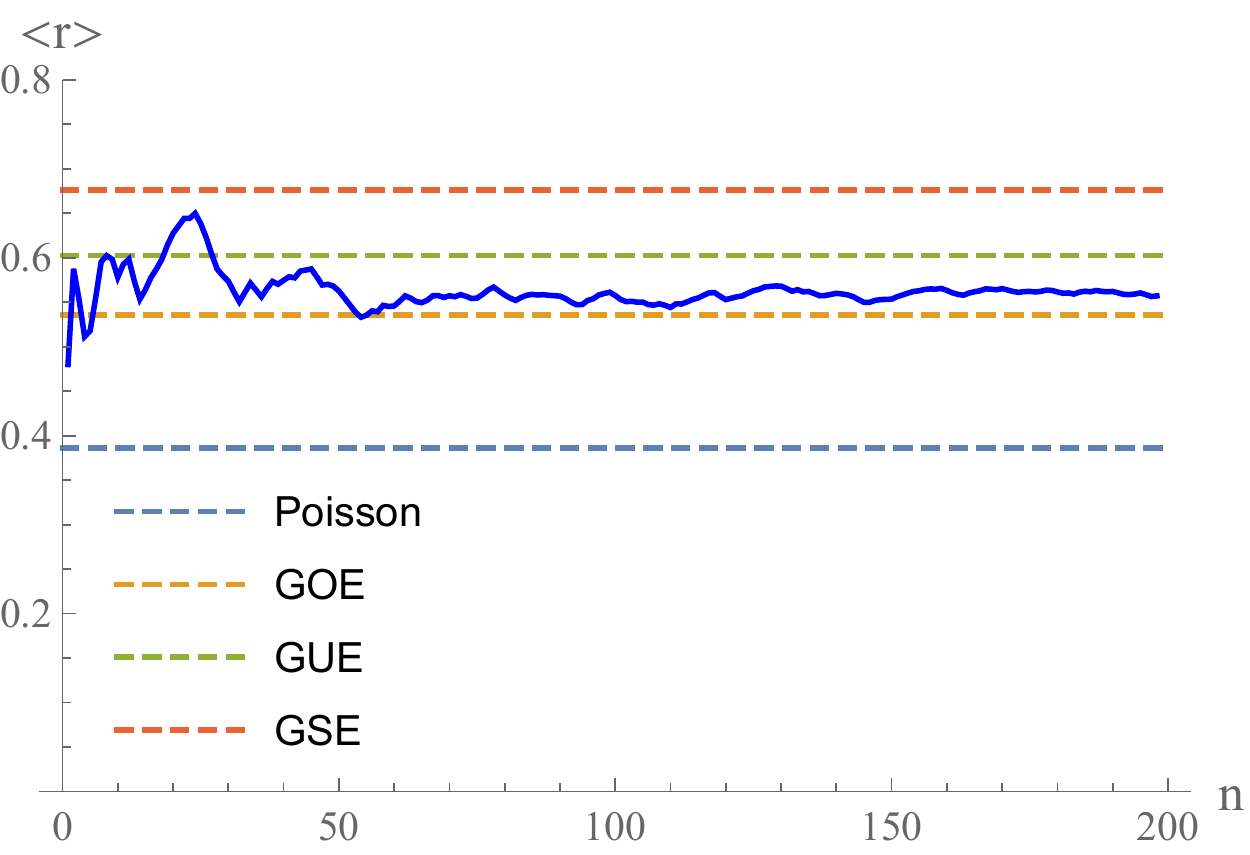}
\end{center}
\end{minipage}
\caption{ {\bf Left:} Plot of the distribution of the adjacent gap ratio $\braket{r_m}$.
{The horizontal axis is the log of the gap ratio $\log r_m$ and the vertical axis is the eigenvalue density.
These are binned by plotting the number of $\log r_m$ in  $[x_0+ m r_0, x_0+(m+1) r_0]$ divided by the total number $M$ for $x_0=0$, $r_0 = 0.5$, $m \in \mathbb{Z}$ and $M = 200$.  }
 {\bf Right:} Plot of the adjacent gap ratio $\braket{r_n}$.
The vertical axis is the average of $r_n$ from $1$ to $N$ levels: $\f{1}{n}\sum_{n=1}^N r_n$.
The horizontal axis is the level $N$ up to which we average. These plots are computed using the numerical results for the first 200 eigenvalues arising from an approximation to the WP metric up to $\mathcal{O}(q^0,(\log(q))^{-60})$.}  
\label{fig:SpectralStatistics1Pn60}
\end{figure}

The probability density of \eqref{eq:defgapratio} diagnoses the random matrix behavior of a quantum system;  it
characterizes the short range correlation of energy levels because it depends on the nearest neighbor spacing of eigenvalues.
By taking the expectation value of the parameter $r$, we can compare our results for the eigenvalues of the WP Laplacian on $\mathcal M_{0,4}$ to the results from RMT.
In Figure  \ref{fig:SpectralStatistics1Pn60}, we compare both  the probability distribution of the adjacent gap ratio $p(r)$ and the average $\braket r$ of our results to that of RMT. Our plots are based on the first 200 eigenvalues computed using an approximation of the WP metric up to $\mathcal{O}(q^0,(\log(q))^{-60})$.\footnote{The plots look essentially identical if we instead include the first 200 eigenvalues computed using an approximation to the WP metric with $q^2$ nonperturbative corrections.}

We note that the statistics of the eigenvalues of the WP Laplacian are far from Poisson; they clearly exhibit the eigenvalue repulsion characteristic of a chaotic system. 
Furthermore, we find that the closest ensemble is the GOE, which reflects the time reversal symmetry of the system.
This is expected because we consider particle motion on the moduli space $\mathcal M_{0,4}$ without the presence of a time reversal symmetry breaking term such as an external magnetic field.
More quantitatively, we compare the average of the $r$ parameter.
In the right plot in Figure \ref{fig:SpectralStatistics1Pn60}, we show the average of the first $n$ eigenvalues as a function of $n$, compared to the expected results from the ensembles presented in Table \ref{table:gapratio}.
As we take more eigenvalues, the expectation value $\braket r$ appears to approach that of the GOE.
We take this as suggestive evidence that the spectral statistics for a particle moving on the moduli space $\mathcal{M}_{0,4}$, with Hamiltonian given by the WP Laplacian, is in the GOE universality class.

\subsection{Level statistics: Spectral form factor}\label{s:SFF}
In this section we analyze the spectral form factor (SFF) of our system.  This is a characterization of the two-point correlations in the eigenvalue spectrum $\langle \rho(E) \rho(E')\rangle$, and can be used to diagnose systems with quantum chaotic behavior. It is sensitive to long-range correlations between the eigenvalues of the system, and can also be used to demonstrate RMT behaviour. 

The (SFF) $g(t; \beta)$\footnote{Strictly speaking, only the infinite temperature ($\beta = 0$) version of the analytically continued partition functions are referred to as spectral form factors in the quantum chaos literature, whereas a finite temperature extension is useful if we consider  systems with infinite dimensional Hilbert spaces. } is defined as the square of an analytically continued thermal partition function  \cite{Haake}, which takes the form,
\ba
g(t;\beta) := |Z(\beta + it)|^2 &=& |\Tr(e^{-\beta H }  e^{-i H t})|^2  \notag \\
&=& \sum_{n,m } e^{-\beta (E_n + E_m )} e^{-i  (E_n- E_m)t}, \label{eq:SFF}
\ea
where $H$ is the Hamiltonian of the system, $\beta$ is the inverse temperature, and $t \geq 0.$ 

\begin{figure}[ht]
\begin{minipage}{0.49\hsize}
\begin{center}
\includegraphics[width=7cm]{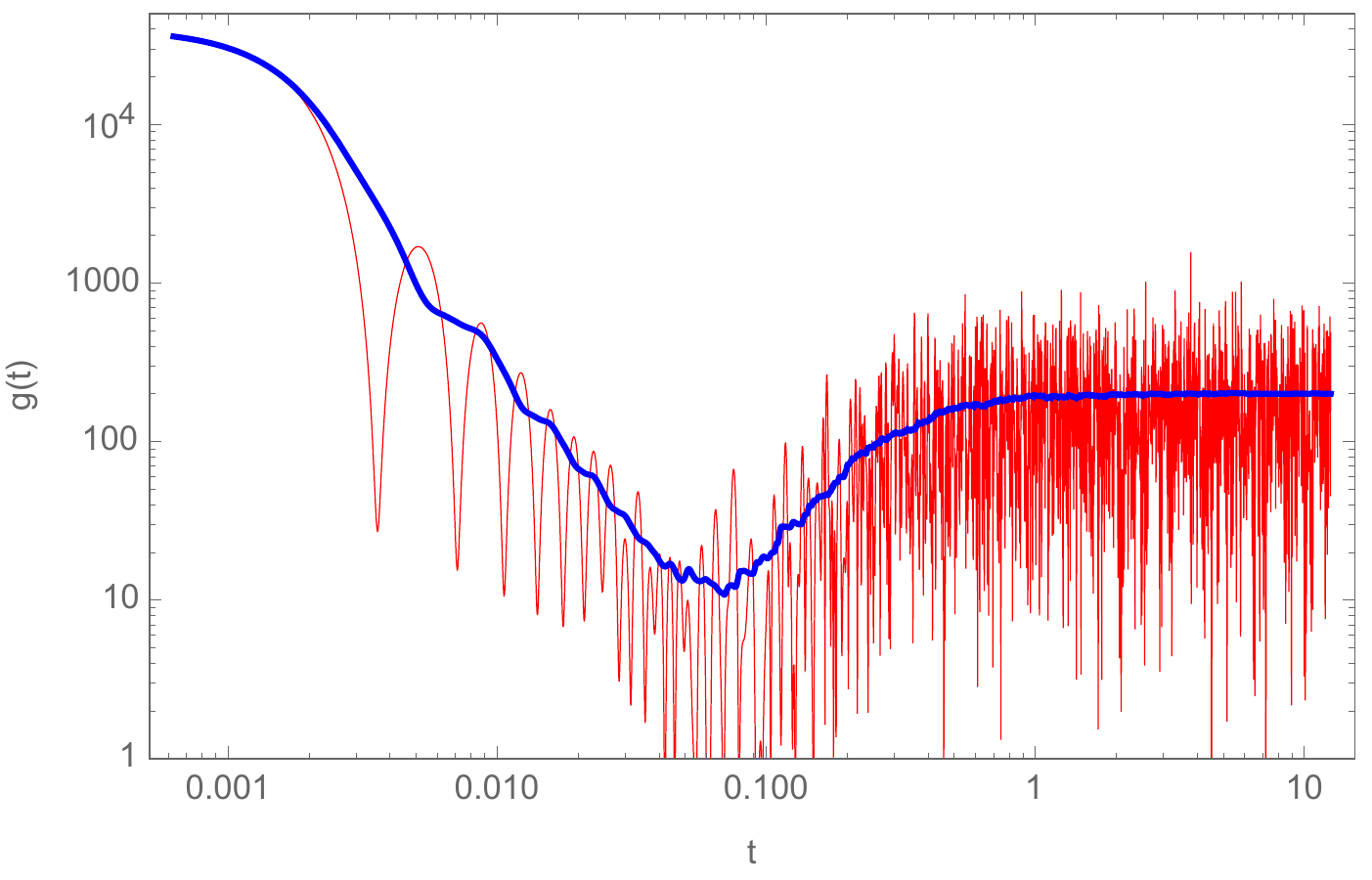}
\end{center}
\end{minipage}
\begin{minipage}{0.49\hsize}
\begin{center}
\includegraphics[width=6.2cm]{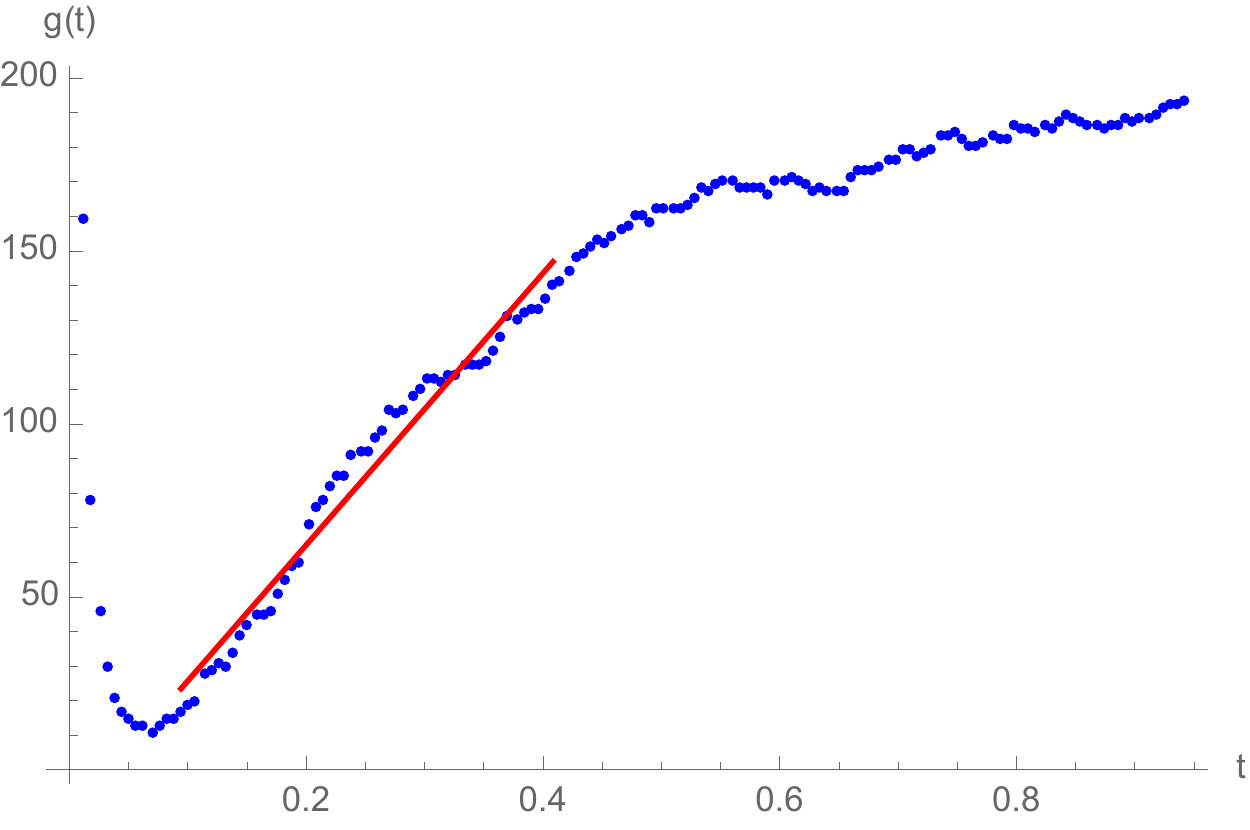}
\end{center}
\end{minipage}
\caption{{\bf Left:} The log-log plot of the spectral form factor.  The blue line indicates the time average.
{\bf Right:} The plot of the spectral form factor (not on log scale).
The red line is the linear fit of region $0.0942 < t < 0.408$.}  
\label{fig:SFFPn60}
\end{figure}

The features of the spectral form factor for systems with random matrix behavior are as follows. Begin by separating the time $t$ into four regimes, following \cite{Cotler:2016fpe}:
\begin{enumerate}
\item At early times $t$, the SFF decays exponentially in $t$. This behavior is often referred to as the slope.
\item At some ``dip time" $t_d$, the SFF reaches a minimum.
\item For $t_{d}<t < t_p$, the SFF increases linearly. This behavior is called the ramp.
\item Finally, at  very late times, $t>t_p$ (the ``plateau time"), the behavior crosses over to become a constant value. This region is referred to as the plateau.
\end{enumerate}

The existence of the ramp characterizes the random matrix behavior of quantum systems.  A linear ramp corresponds to long-distance level repulsion in the energy eigenvalues, with  $\langle \rho(E) \rho(E')\rangle\propto\frac{1}{(E-E')^2}$.
The early time decay is related to thermal relaxation and in dual theories of gravity is related to quasi-normal modes of black holes \cite{Cotler:2016fpe,Horowitz:1999jd}.
The plateau at late times is a reflection of the discreteness of the spectrum.  In particular, the long-time average of the spectral form factor approaches  
\be
\lim_{T \to \infty}\f{1}{T}\int _0^T dt |Z(\beta + i t)|^2 = \sum _n N_E e^{-2\beta E_n}
\ee
at late time, where $N_E$ is the degeneracy in the number of states at energy $E$.
If there are no degeneracies the late time value is given by the partition function $Z(2\beta)$.

The SFF of a chaotic theory typically exhibits erratic oscillations after the dip time, unless some sort of averaging is employed \cite{1997PhRvL..78.2280P}.\footnote{For a Hamiltonian which is drawn from an ensemble (such as SYK) one can implement this averaging simply by averaging over draws from the ensemble.  In the present case we have a single Hamiltonian so must employ a different type of averaging.} 
Here we employ the time average \cite{Balasubramanian:2016ids}
\be
g_{\text{t-ave}} (t;\beta) := \f{1}{at}\int_{t - \f{at}{2}}^{t+ \f{at}{2}} |Z(\beta + i t)|^2 dt,
\ee 
with $a=1$.
Note that the window over which we average grows linearly in time, so is a constant in the log-log plot.

The spectral form factors for the GOE, GUE and GSE ensembles of $N\times N$ matrices take the form \cite{Liu:2018hlr},
\begin{align}
&|Z_{\text{GOE}}(it)|^2 = Z_{\text{dip}}^2 (t) +  N \times
\begin{cases}
t - \f{t}{2} \log \big(1 + t \big) \qquad t \le 2  \\
2  - \f{t}{2} \log \big( \f{t+1}{t-1} \big)\qquad t > 2  \\
\end{cases}  \notag \\
& |Z_{\text{GUE}}(it)|^2 = Z_{\text{dip}}^2 (t) +  N \times
\begin{cases}
\f{t}{2}  \qquad t \le 2  \\
1 \qquad t > 2  \\
\end{cases}  \notag \\
& |Z_{\text{GSE}}(it)|^2 = Z_{\text{dip}}^2 (t) +  N \times
\begin{cases}
\f{t}{4} - \f{t}{8} \log \big| 1 - \f{t}{2} \big|  \qquad t \le 4  \\
1 \qquad t > 4  \\
\end{cases}  \notag \\
\end{align}
where 
\be
Z_{\text{dip}}^2(t) =\f{J_1(2N t)^2}{t^2},
\ee
and $J_1$ is the Bessel function of the first kind.

In this paper, we only study the spectrum of the symmetric sector of eigenfunctions, and thus 
our SFF is defined as,
\be
g_{\text{sym}}(t;\beta) := |Z(\beta + it)|^2 = |\Tr( P_{\text{sym}}e^{-\beta H }  e^{-i H t})|^2  ,
\ee
where $P_{\text{sym}}$ is the projection onto the symmetric sector defined by  \eqref{eq:S3sym} and \eqref{eq:conjugtion}.
Furthermore, we can only compute a finite number of eigenvalues in the spectrum numerically, so we truncate the sum over the spectrum as,
\be
g_{\text{sym}}^{\text{truncate}-N}(t;\beta) := \sum_{i,j=1}^N e^{-\beta(E_i + E_j)} e^{-i (E_i - E_j)t}. 
\ee
Here $i,j$ label the spectrum of eigenvalues in the symmetric sector. In Figure  \ref{fig:SFFPn60}, we plot the result for the SFF $g_{\text{sym}}^{\text{truncate}-200}(t;\beta)$ using the first 200 numerically-computed symmetric sector eigenvalues. Even though we truncate to only 200 eigenvalues, we are nevertheless clearly able to  see a ramp characteristic of long distance eigenvalue repulsion. The ramp region is well fit by a linear function.
This is further evidence that two-point statistics of eigenvalues exhibits features characteristic of quantum chaos.

\section*{Acknowledgements}

We thank J. Chandra, S. Collier, K. Colville, T. Hartman, K. Namjou, S. Shenker, and H. Verlinde for helpful conversations.  
The work of S.M.H. is supported by the National Science and Engineering Council of Canada and the Canada Research Chairs program.
Research of AM is supported in part by the Simons Foundation Grant No. 385602 and the Natural Sciences and Engineering Research Council of Canada (NSERC), funding reference number
SAPIN/00047-2020.  
TN is supported by  MEXT KAKENHI Grant-in-Aid for Transformative Research Areas A ``Extreme Universe'' Grant Number 22H05248.

\appendix 
\section{Numerical results for the eigenvalues}\label{app:eigenvalues}

\subsection{Numerical Data}\label{ap:nd}

In this section, we present the numerical data for the eigenvalues of the Laplacian on the moduli space $\mathcal{M}_{0,4}$.
As discussed in the main text, for simplicity we consider only the eigenfunctions that obey \eqref{eq:S3sym} and \eqref{eq:conjugtion}.
The results are shown in tables \ref{table:SpectrumP65uv10} and \ref{table:SpectrumNP30uv100}.  Our methods can easily be  adapted to compute the  spectrum up to arbitrarily high energy levels.

\begin{table}[htb]
  \begin{center}

    \begin{tabular}{cccccccc}
    \hline
  24.004 & 39.8792 & 56.4147 & 69.6923 & 75.6385 & 92.8046 & 102.244 & 114.388 \\
 128.198 & 137.274 & 142.367 & 155.89 & 166.06 & 181.288 & 185.131 & 199.328 \\
 209.627 & 217.81 & 228.69 & 237.372 & 245.334 & 254.32 & 265.377 & 278.709 \\
 287.302 & 298.239 & 302.096 & 320.042 & 321.723 & 335.921 & 341.145 & 354.566 \\
 357.093 & 369.811 & 381.324 & 391.954 & 395.11 & 406.473 & 418.95 & 433.165 \\
 439.53 & 444.306 & 450.738 & 462.997 & 474.166 & 481.323 & 492.444 & 494.414 \\
 509.797 & 519.197 & 523.611 & 538.689 & 541.022 & 558.767 & 560.818 & 570.013 \\
 584.012 & 595.434 & 600.936 & 606.644 & 618.73 & 625.725 & 631.78 & 638.204 \\
 654.643 & 659.605 & 670.042 & 677.64 & 686.185 & 701.319 & 707.672 & 712.033 \\
 721.206 & 727.92 & 744.16 & 750.174 & 760.372 & 769.299 & 778.799 & 790.252 \\
 792.521 & 803.731 & 806.792 & 815.748 & 826.515 & 834.676 & 847.554 & 853.963 \\
 864.855 & 870.227 & 880.657 & 885.938 & 903.13 & 906.36 & 918.201 & 924.741 \\
 \hline
\end{tabular}  \caption{First $96$ eigenvalues of the Laplacian on $\mathcal{M}_{0,4}$ at order $\mathcal{O}(q^0,(\log(q))^{65})$ (leading order in the conformal block expansion). 
    The area of each discrete triangle in the fundamental domain with respect to the flat metric is taken to be smaller than $10^{-5}$. }\label{table:SpectrumP65uv10}
\end{center}
\end{table}

\begin{table}[htb]
\begin{center}

    \begin{tabular}{cccccccc}
    \hline
   24.0024 & 39.8759 & 56.4075 & 69.6792 & 75.6396 & 92.7882 & 102.233 & 114.381 \\
 128.155 & 137.265 & 142.374 & 155.857 & 166.042 & 181.265 & 185.093 & 199.308 \\
 209.633 & 217.727 & 228.651 & 237.369 & 245.359 & 254.249 & 265.327 & 278.66 \\
 287.275 & 298.21 & 302.043 & 320.044 & 321.644 & 335.852 & 341.113 & 354.564 \\
 357.033 & 369.773 & 381.301 & 391.801 & 395.131 & 406.401 & 418.892 & 433.118 \\
 439.428 & 444.291 & 450.719 & 462.908 & 474.116 & 481.308 & 492.347 & 494.366 \\
 509.72 & 519.182 & 523.601 & 538.573 & 540.922 & 558.692 & 560.758 & 569.982 \\
 583.947 & 595.34 & 600.838 & 606.606 & 618.659 & 625.728 & 631.74 & 638.133 \\
 654.622 & 659.505 & 670.045 & 677.442 & 686.093 & 701.272 & 707.663 & 711.971 \\
 721.259 & 727.768 & 743.975 & 750.081 & 760.377 & 769.144 & 778.855 & 790.136 \\
 792.511 & 803.755 & 806.626 & 815.756 & 826.47 & 834.588 & 847.531 & 853.984 \\
 864.801 & 870.031 & 880.593 & 885.793 & 903.137 & 906.308 & 918.223 & 924.752 \\
 \hline
\end{tabular}\caption{First $100$ eigenvalues of the Laplacian on $\mathcal{M}_{0,4}$ at order $\mathcal{O}(q^2,(\log(q))^{30})$ (including $O(q^2)$ corrections in the conformal block expansion) . 
    The area of each discrete triangle  in the fundamental domain with respect to the flat metric is taken to be smaller than $10^{-4}$. }\label{table:SpectrumNP30uv100}
\end{center}
\end{table}

\subsection{On the accuracy of the eigenvalues}
We now estimate the errors on eigenvalues that we numerically compute.
These will depend on both the expansion in the coordinate $q$ and the mesh size $\Delta A$ used when we discretize moduli space.

First we consider the finite $\Delta A$ effects.
We are studying the eigenvalue problem defined by,
\be
e^{-2\phi(x_1,x_2)} (\partial_{x_1}^2 + \partial_{x_2}^2) \psi(x_1,x_2) = E_n\psi(x_1,x_2),
\ee 
where $x_1 = \text{Im} \tau$ and $x_2 = \text{Re} \tau$.
Because the lattice spacing is given by $\s{\Delta A}$, we estimate that the cut off of the ``momentum" $\partial_{x_i}$ is given by $1/\s{\Delta A}$. 
To get a tighter bound, we replace $e^{2\phi(x_1,x_2)}$ by $\text{max} ( e^{2\phi(x_1,x_2)})$.
Then, we expect that the cutoff of the energy is
\be
E_n \le \f{1}{\text{max} ( e^{2\phi(x_1,x_2)})} \f{1}{\Delta A }.
\ee
Using  Weyl's law $N(E) \approx \f{A}{4\pi} E$, we can trust the numerical results for eigenvalues which satisfy 
\be
N(E) \ll \f{1}{4\pi}  \f{1}{\text{max} ( e^{2\phi(x_1,x_2)})} \f{A}{\Delta A }.
\ee
For the sector of eigenfunctions we study on $\mathcal M_{0,4}$, $\text{max} ( e^{2\phi(x_1,x_2)}) = e^{2\phi(\f{1}{2},\f{\s{3}}{2})} \approx  0.8$ and $\f{A}{4\pi} = 0.13$.
Therefore, for example, when  $\Delta A = 10^{-4}$, the bound on the eigenvalue is 
\be
N(E) \ll 1600.
\ee

\begin{figure}[ht]
\begin{center}
\includegraphics[width=6.5cm]{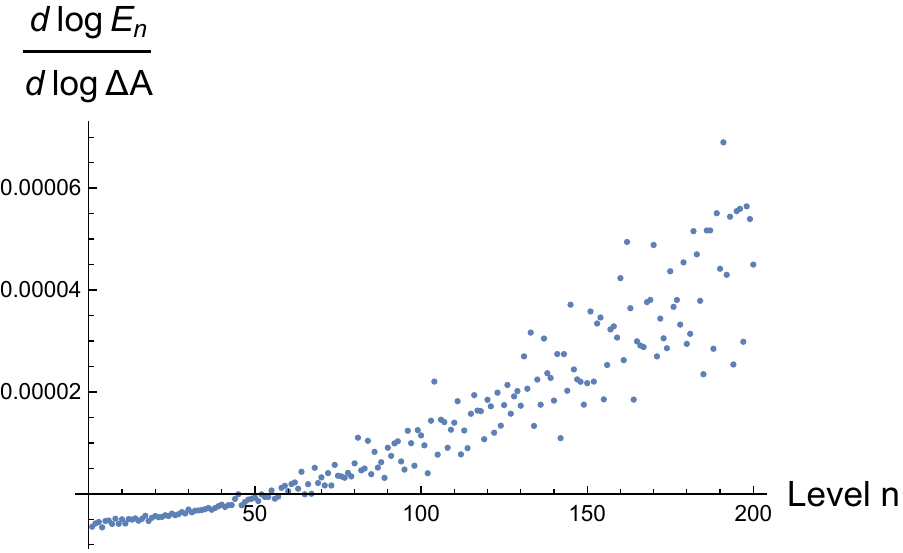}
\includegraphics[width=6.5cm]{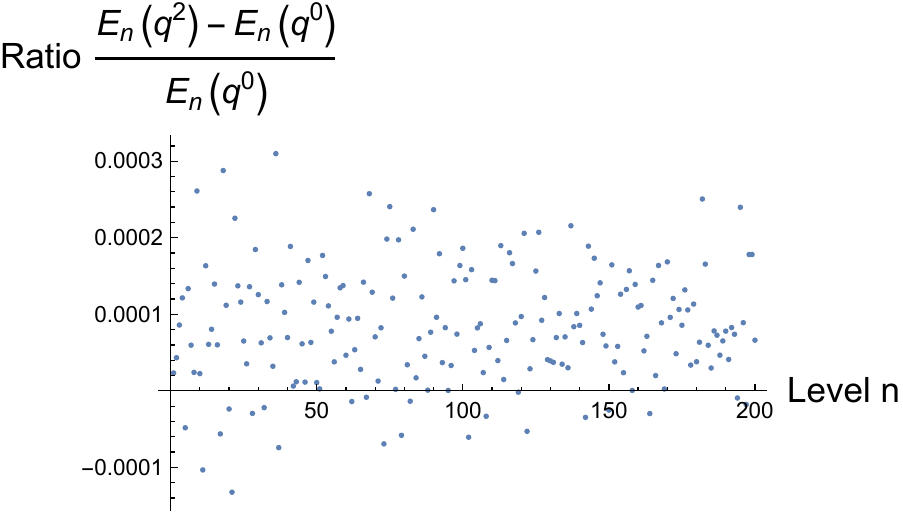}
\caption{{\bf Left:} The plot of $\f{d\log E_n}{d\log \Delta A}$ for each eigenvalues $E_n$ for $ n = 1 \sim 200$ when the area $\Delta A$ of  each mesh is smaller than $10^{-5}$.
To numerically calculate, we change the mesh size from $5.0 \times 10^{-5}$ to $1.0 \times 10^{-5}$ and use the approximate $\f{d \log E_n}{d\log \Delta A}$ by the ratio of differences.
{\bf Right:} The shifts of eigenvalues when we include the $\mathcal{O}(q^2)$ corrections to the metric. 
$E_n(q^2)$ means the $n$-th eigenvalue at the order $\mathcal{O}(q^2, (\log |q|)^{-30})$ whereas $E_n(q^0)$ is that at the order  $\mathcal{O}(q^0, (\log |q|)^{-30})$.
}  
\label{fig:SpectrumStability}
\end{center}
\end{figure}

To visualize how the spectrum are stable under the change of mesh size or non-perturbative corrections,
we show the shifts of eigenvalues in figure \ref{fig:SpectrumStability}.

Next, we consider the effect of the finite truncation at $\mathcal{O}((\log |q|)^{-n},q^m)$.  As $n$ is increased, the new subdleading terms in $g_{q{\bar q}}$ are suppressed by powers of $\left(\log \frac{2^8}{q{\bar q}}\right)^{-1}$, which is quite small in the fundamental domain for $x$.  For example at $x=\frac{1}{2}$ (on the boundary of the fundamental domain, where the errors will be the largest) we have $\left(\log \frac{2^8}{q{\bar q}}\right)^{-1}\approx 0.08$.  Similarly, as $m$ is increased subleading terms will be introduced which are suppressed by powers of $q^2$, which at $x=\frac{1}{2}$ is $q^2\approx 0.002$.  This explains the results in Figure \ref{fig:Volumedif}, where we see roughly 20-30 terms in the perturbative expansion (i.e. expansion in $n$) are needed before the non-perturbative expansion (the expansion in $m$) becomes important.  It would be interesting to investigate the size of the error terms more quantitatively, which would require a study of the growth of the numerical coefficients which appear in this expansion; this may be possible using the Zamolodchikov recursion recursion relations, but we will leave this for future work.  It is, however, possible to check explicitly that the numerical results for the eigenvalues change only by small amounts as subleading terms are introduced; this is clear by comparing tables in \ref{ap:nd}.  In particular, by introducing an additional non-perturbative term in our approximation, the values of the eigenvalues only change by roughly one part in $10^{-4}$.

\clearpage
\bibliography{M04refs}
\end{document}